\documentclass[journal,twocolumn]{IEEEtran}
\usepackage{}
\usepackage{graphicx}
\usepackage{epstopdf}
\usepackage{amssymb}
\usepackage{amsfonts}
\usepackage{amsmath}
\usepackage{algorithm}
\usepackage{algorithmic}
\usepackage{subeqnarray}
\usepackage{cases}
\usepackage{bm}
\usepackage{color}
\usepackage{subfigure,amsmath,amssymb,cite}
\usepackage{float}
\ifCLASSINFOpdf
\else
\fi
\begin{document}

%\title{Coordinated Hybrid Precoding and Phase Shift for Intelligent Reconfigurable Surface-aided Directional Modulation Network}
\title{Joint Beamforming and Phase Shift Design for Hybrid-IRS-and-UAV-aided Directional Modulation Network}
%\title{Hybrid RIS}
\author{Rongen Dong, Hangjia He, Feng Shu, Qi Zhang, Riqing Chen, \\
Shihao Yan, and Jiangzhou Wang,\emph{ Fellow, IEEE}

%\thanks{Copyright (c) 2015 IEEE. Personal use of this material is permitted. However, permission to use this material for any other purposes must be obtained from the IEEE by sending a request to pubs-permissions@ieee.org.}
\thanks{This work was supported in part by the National Natural Science Foundation of China (Nos.U22A2002, 62071234, and 61972093), the Major Science and Technology plan of Hainan Province under Grant ZDKJ2021022, the Scientific Research Fund Project of Hainan University under Grant KYQD(ZR)-21008, and the Fujian University Industry University Research Joint Innovation Project (No.2022H6006).}
%(Corresponding author: Feng Shu).
%the Innovative Research Project of Postgraduates in Hainan Province (Qhyb2022-83)% +创新项目号
%%\thanks{Tong Shen,~Jin Wang,~and Feng Shu are with the School of Electronic and Optical Engineering, Nanjing University of Science and Technology, 210094, CHINA.. }
\thanks{Rongen Dong, Feng Shu, and Qi Zhang are with the School of Information and Communication Engineering, Hainan University, Haikou, 570228, China (Email: shufeng0101@163.com).}
\thanks{Hangjia He is with the School of Electronic and Optical Engineering, Nanjing University of Science and Technology, Nanjing, 210094, China.}
\thanks{Riqing Chen is with the Digital Fujian Institute of Big Data for Agriculture, Fujian Agriculture and Forestry University, Fuzhou 350002, China (Email: riqing.chen@fafu.edu.cn).}
\thanks{Shihao Yan is with the School of Science and Security Research Institute, Edith Cowan University, Perth, WA 6027, Australia (e-mail: s.yan@ecu.edu.au).}
\thanks{Jiangzhou Wang is with the School of Engineering, University of Kent, Canterbury CT2 7NT, U.K. (Email: {j.z.wang}@kent.ac.uk).}

% <-this % stops a space
%
}

\maketitle

\begin{abstract}
Recently, intelligent reflecting surface (IRS) and unmanned aerial vehicle (UAV) have been introduced into wireless communication systems to enhance the performance of air-ground transmission. To make a good balance between performance, cost, and power consumption, a hybrid-IRS-and-UAV-assisted directional modulation (DM) network is investigated in this paper, where the hybrid IRS consists of passive and active reflecting elements. To maximize the achievable rate, three optimization algorithms, called maximum signal-to-noise ratio (SNR)-fractional programming (FP) (Max-SNR-FP), maximum SNR-equal amplitude reflecting (EAR) (Max-SNR-EAR), and maximum SNR-majorization-minimization (MM) (Max-SNR-MM), are proposed to jointly design the beamforming vector and phase shift matrix (PSM) of hybrid IRS by alternately optimizing one and giving another. The Max-SNR-FP method employs the successive convex approximation and FP methods to derive the beamforming vector and hybrid IRS PSM. The Max-SNR-EAR method adopts the maximum signal-to-leakage-noise ratio method and the criteria of phase alignment and EAR to design them. In addition, the Max-SNR-MM method utilizes the MM criterion to derive the IRS PSM. Simulation results show that the rates harvested by the proposed three methods are slightly lower than those of active IRS with higher power consumption, which are 35 percent higher than those of no IRS and random phase IRS, while passive IRS achieves only about 17 percent rate gain over the latter.
Moreover, compared to Max-SNR-FP, the proposed Max-SNR-EAR and Max-SNR-MM methods make an obvious complexity degradation at the price of a slight performance loss.

\end{abstract}
\section{Introduction}
Nowadays, wireless networks serve a wide range of civilian and military applications and have become an essential part of our routine\cite{Wu2018A, Zheng2013Improving}. Improving the performance of wireless communication has become a popular research topic.
Unmanned aerial vehicle (UAV), thanks to its low cost, high flexibility and high probability of line-of-propagation (LoP) links, has become an attractive means of improving air-ground transmission quality \cite{Zeng2016Wireless}. In general, UAVs can not only act as base stations\cite{Yin2020Resource}, relays\cite{Zhang2018Joint}, etc, to transmit or forward signals, but also for data collection\cite{Zhou2021UAV} and positioning\cite{Wen20233D}. Currently, UAV transmission technology has been widely explored. For instance, the authors in \cite{Azari2020UAV} considered a cellular network deployment in which UAV-to-UAV launch-receive pairs used the identical spectrum as the uplink of cellular ground users, and the performances of underlay and overlay spectrum sharing mechanisms were analyzed and compared. In \cite{Zeng2021Trajectory}, to enhance the throughput of a single-cell multi-user orthogonal frequency division multiple access network with single UAV while guaranteeing the user fairness, an efficient method was proposed that outperformed the random and cellular schemes in terms of user fairness and sum rate. UAV-enabled relay under malicious jamming was considered in \cite{Wu2021UAV}, and the successive convex approximation (SCA) algorithm was employed to maximize the end-to-end throughput. However, UAV network also presents some challenges that affect their performance. For instance, the LoP links may be blocked, UAV forwarding signal increases power consumption and affects the endurance of UAV, etc.

As an effective solution to the above mentioned problems, intelligent reflecting surface (IRS) has been investigated as an intelligent and reconfigurable paradigm for future wireless communications\cite{Wu2019Intelligent}. IRS is a plane composed of lots of reflective elements, which can intelligently tune the amplitude and phase of the incident signal to reconfigure the wireless transmission environment and has proven to be an energy and cost-efficient tool for enhancing
the performance of the wireless network. Driven by these advantages, IRS-aided UAV networks have been widely investigated.
A network that employed UAV and IRS to support terahertz communication was considered in \cite{Pan2021UAV}, and an iteration method was proposed to maximize the minimum average achievable rate among all users.
In \cite{Su2022Spectrum}, two approaches were proposed to maximize the spectrum and energy effectiveness of the IRS-aided UAV network by jointly deriving the UAV trajectory, active and passive beamforming, and the proposed methods yielded a better performance compared to the baselines. A secure IRS-aided UAV network was investigated in \cite{Fang2021Joint}, and a SCA scheme was proposed to maximize secrecy rate (SR).
To maximize the average SR of the IRS-aided secure UAV system in \cite{Pang2022IRS}, the fractional programming and SCA methods were applied to  deal with the non-convex optimization problem. The authors considered a symbiotic UAV-aided IRS radio network in \cite{Hua2021UAV}, a relaxation-based scheme was proposed to minimize the weighted sum bit error rate of all IRSs.

Directional modulation (DM), which has been shown to significantly boost the rate of wireless communication system, has evolved into a useful strategy for fifth-generation millimeter-wave communication system\cite{Cheng2021Physical, Wang2018Hybrid, Nusenu2019Development}. The DM is capable of guiding standard baseband symbols to the desired direction while distorting the signal constellation diagram outside that direction\cite{Qiu2020Multi}. The design of DM synthesis is mainly carried out at the radio frequency frontend or baseband. For example, in \cite{Daly2009Directional}, the signal was generated in a predetermined direction through tuning the phase of each antenna element at the radio frequency frontend. The authors in \cite{Daly2010Beamsteering} sketched a process for determining that how to switch the antenna elements to send signals only in a given direction, and the DM array can transmit signals over a narrower beamwidth compared to the traditional reconfigurable array. In \cite{Shu2016Robust}, based on the maximizing signal-to-artificial noise (AN) ratio and maximizing signal-to-leakage-noise ratio approaches designed at baseband, the AN projection matrix and precoder vector were obtained to maximize the SR of a multi-beam DM network. An AN-aided zero-forcing synthesis scheme that achieved the dynamic characteristics of multi-beam DM through random varying of AN vector was proposed in \cite{Xie2018Artificial}, which was easier and more effective to implement than the traditional dynamic multi-beam DM synthesis methods in wireless network with a certain performance loss. In \cite{Teng2022Low}, the authors investigated a DM system with malicious jamming, and proposed three receive beamforming algorithms to boost the SR.

In particular, conventional DM networks can only transmit single bit stream, while the emergence of IRS made it possible for DM networks to transmit multiple bit streams. IRS with energy and cost-efficient can be employed to create friendly and controllable multipaths to transmit two bit stream or increase rate to enhance the performance of DM systems. For example,
%Recently, IRS-assisted DM systems have also been investigated.
in \cite{ShuEnhanced2021}, to transmit two bit streams from Alice to Bob and maximize the secure rate of IRS-assisted DM network, the precoder vectors and phase shift matrix (PSM) of IRS were jointly devised by a high-performance general alternating iterative and low-complexity null-space projection algorithms.  For maximizing the receive power sum of IRS-assisted DM system, the authors in \cite{Dong2022Low} proposed the general alternating optimization and zero-forcing methods to jointly design the PSM at IRS and receive beamforming vectors at user. In \cite{Chen2022Artificial}, aiming to maximize the security rate of IRS-assisted multiple-input single-output (MISO) DM network, the semi-definite relaxation algorithm was utilized to derive the precoding and IRS PSM when the location information of the eavesdropper was available. Thanks to the combination with IRS, the security rate performance of them has been significantly boosted.

%Moreover,IRS is also available to assist various wireless communication scenarios: unmanned aerial vehicle (UAV) communication , cell wireless communication \cite{Wu2019Intelligent, Pan2020Multicell}, non-orthogonal multiple access network\cite{Zheng2020Intelligent, Fang2020Energy}, etc.

However, all the above work was done based on fully passive IRS, and a satisfactory achievable rate of the system may not be ensured due to the effect of ``double fading'' in the cascaded channels. To effectively combat this effect and enhance the performance of the passive IRS-aided wireless communication network, the fully active IRS has been investigated recently \cite{Zhang2021Active, Liu2022Active, Ren2023Transmission, Dong2022Active, Lv2023RIS}. Their simulation results showed that active IRS achieved significant performance improvements compared to passive IRS.
%In \cite{Long2021Active}, to maximize signal-to-noise ratio of the active IRS-assisted single input multiple output network, the linear minimum-mean-square-error criterion and SCA method were adopted to design the receive beamforming and reflecting coefficient matrix, respectively. In \cite{You2021Wireless}, the downlink and uplink communication networks were considered separately, and the placement of active or passive IRS was investigated to maximize the achievable rate. An active IRS-assisted secure wireless network was considered in \cite{Dong2022Active}, and an alternating optimization method was proposed to achieve higher secrecy performance than passive IRS and without IRS. In \cite{Lv2023RIS}, the authors aimed to minimize the transmit power in an active IRS-aided MISO network, and proposed a penalty-based alternating minimization method to derive the precoder and IRS PSM, the simulation results showed that active IRS outperforms passive IRS in terms of energy efficiency. To alleviate the hardware cost and energy consumption caused by the large number of active elements in fully-connected active IRS, a sub-connected active IRS system was investigated in \cite{Zhu2022Joint}, and four methods were employed to tackle the sum rate maximization and power minimization problems.
However, the higher rate achieved by active IRS comes at the price of high hardware cost and power consumption \cite{Nguyen2022Hybrid2}. To overcome the limitations of fully passive and active IRSs, a hybrid active-passive IRS was proposed \cite{Nguyen2022Hybrid}. The main idea of the hybrid IRS is employing some active elements to substitute the one of the passive IRS, these active elements with signal amplification of hybrid IRS can effectively make up for the cascade path loss (PL) and increase the achievable rate\cite{Nguyen2021Spectral}. In \cite{Ngo2021Low}, to explore the potential of hybrid relay-IRS in helping single-user mobile edge computing systems for computational offloading, an efficient scheme was proposed to design the received beamforming, IRS reflection coefficients and computational parameters, and the latency was dramatically reduced with the help of IRS. In \cite{Hu2021Hybrid}, a hybrid IRS-assisted covert transmission network was proposed to boost the performance of traditional covert communication network, and an alternate algorithm with closed-form expression was derived to obtain the transmit power and IRS PSM. A hybrid IRS-assisted integrated sensing and communication network with multiple users and targets was investigated in \cite{Sankar2022Beamforming}, aimed at maximizing the worst-case target illumination power, an approach was proposed to design the precoding and IRS coefficients.

In fact, the IRS needs to be installed on the surface of the object. However, when the installation plane is absent and/or emergency communication is called for, such as disaster relief, mounting the IRS on the UAV is an effective way to solve this challenge. In this way, not only can signal enhancement be achieved, but also the position of the IRS can be flexibly adjusted, and the signal coverage may be extended.
Given the benefits of hybrid IRS-assisted wireless networks performance, it is a reasonable choice to combine them with the conventional DM networks to balance the performance and power consumption. So far, as far as the authors know, the hybrid-IRS-and-UAV-aided DM system have not been investigated yet. In this article, we employ the hybrid IRS to further enhance the performance of passive IRS-aided DM network. The main contributions of this work are summarized as follows:

\begin{enumerate}
\item To make a good balance between performance, cost, and power consumption, a hybrid-IRS-and-UAV-aided DM system model is proposed. Aiming at maximizing the achievable rate, the optimization problem of maximizing the signal-to-noise ratio (SNR) is established,
%    This problem is non-convex and difficult to solve directly due to the hybrid IRS power constraint, the constant modulus phase shift constraint, and the coupled optimization variables.
    and the maximum SNR-fractional programming (FP) (Max-SNR-FP) method is proposed to jointly optimize the transmit beamforming vector and hybrid IRS PSM by solving one and giving another. In this scheme, the beamforming vector and passive IRS PSM are obtained by the successive convex approximation algorithm, and the active IRS PSM is computed by the FP method.

\item Given the high computational complexity of the Max-SNR-FP scheme, a low-complexity method, named maximum SNR-equal amplitude reflecting (EAR) (Max-SNR-EAR), is proposed. In this method, by utilizing the maximum signal-to-leakage-noise ratio (SLNR) criterion, the beamforming vector is obtained. In addition, the phases of the passive and active IRS phase shift matries are computed based on the criteria of phase alignment, while the amplitude of the active IRS PSM is obtained by the EAR criterion.

\item Given that the passive and active IRS phase shift matrices of the proposed Max-SNR-FP and Max-SNR-EAR methods are designed separately, to investigate the effect of joint design phase shift matrix on system performance improvement, a low-complexity algorithm, called Max-SNR-MM, is proposed to maximize the SNR. The majorization-minimization (MM) criterion is employed to solve the hybrid IRS phase shift matrix. From the simulation results, it is clear that the achievable rates harvested by the proposed three methods are higher than those of without IRS, random phase IRS, and passive IRS. In addition, when the number of hybrid IRS phase shift elements tends to be large, the difference in achievable rates between these three proposed methods is trivial.

\end{enumerate}

The remainder of our work is organized as follows. In Section \ref{s1}, we describe the system model of hybrid-IRS-and-UAV-aided DM network.
Section \ref{s2} presents the Max-SNR-FP scheme.
The Max-SNR-EAR scheme is described in Section \ref{s3}.
The Max-SNR-MM scheme is given in Section \ref{s4}.
We show the numerical simulation results in Section \ref{s5}. In Section \ref{s6}, the conclusions are drawn.

{\bf Notations:} in this article, the vectors and matrices are shown in boldface lowercase and uppercase letters, respectively. Symbols $(\cdot)^*$, $(\cdot)^T$, $(\cdot)^H$, Tr$(\cdot)$, $\Re\{\cdot\}$, $\lambda_{\text{max}}\{\cdot\}$, $\text{diag}\{\cdot\}$, and $\text{blkdiag}\{\cdot\}$ stand for the conjugate, transpose, conjugate transpose, trace, real part, maximum eigenvalue of the matrix, diagonal, and block diagonal matrix operations, respectively. The sign $|\cdot|$ refers to the scalar's absolute value or the matrix's determinant. The notations $\textbf{I}_N$ and $\mathbb{C}^{N\times N}$ stand for the identity matrix and complex-valued matrix space of $N\times N$, respectively.

\section{system model}\label{s1}
\begin{figure}[htbp]
\centering
\includegraphics[width=0.45\textwidth]{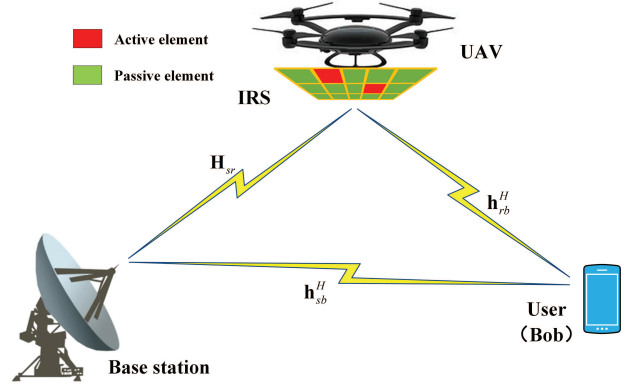}\\
\caption{System diagram of hybrid-IRS-and-UAV-aided DM network.}\label{model}
\end{figure}

%\subsection{HAP-RIS-aided Directional Modulation System}
As indicated in Fig.~\ref{model}, a hybrid-IRS-and-UAV-aided DM network is taken into account, where the IRS is installed on the UAV. Assuming that the UAV operates at a sufficient altitude, and all channels are the line-of-sight channels. There are a base station (BS) with $N$ antennas and a user Bob with single antenna. The hybrid IRS is equipped with $M$ elements, which consists of $M_a$ active and $M_p$ passive IRS reflecting elements ($1\leq M_a\leq M_p, M=M_a+M_p$). It is assumed that the active elements enable adjust both the amplitude and phase while the passive ones only tunes the phase of the incident signal. The signals reflected more than or equal to twice on the hybrid IRS are negligible due to the severe PL\cite{Pan2020Multicell}. We suppose that all the channel state information is completely accessible owing to the channel estimation\cite{Wang2020Channel}.

% bmeps -c model.jpg model.eps(convert a to b)
%Similar to \cite{Shen2022Multi}, we establish a coordinate system with BS as the origin $(0, 0, 0)$, and the location of the IRS and Bob are $(x_r, y_r, z_r)$ and $(x_b, y_b, z_b)$, respectively. In accordance with the trigonometric function operations, the azimuth angles and pitch angles are
%\begin{align}
%&\vartheta_{si}=\text{arctan}\frac{y_i}{x_i}, ~i=r, b,\\
%&\varphi_{si}=\text{arctan}\frac{z_i}{\sqrt{x^2_i+y^2_i}}, ~i=r, b,\\
%&\vartheta_{rb}=\text{arctan}\frac{y_r-y_b}{x_b-x_r},\\
%&\varphi_{rb}=\text{arctan}\frac{z_r-z_b}{\sqrt{(x_r-x_b)^2+(y_r-y_b)^2}}.
%\end{align}
%Accordingly, the distances from BS to IRS, BS to Bob, and IRS to Bob are
%\begin{align}
%&d_{sr}=\sqrt{x^2_r+y^2_r+z^2_r},\\
%&d_{sb}=\sqrt{x^2_b+y^2_b+z^2_b},\\
%&d_{rb}=\sqrt{(x_r-x_b)^2+(y_r-y_b)^2+(z_r-z_b)^2}.
%\end{align}
%The phase of the received signal from the $n$-th antenna of the transmitter is expressed as
%\begin{align}
%&\varpi_n=2\pi\frac{d-(n-1)d_A\text{cos}\gamma_A}{\lambda}\\
%&\gamma_A=\text{arccos}(\text{cos}\varphi_{si}\text{cos}\vartheta_{si})
%\end{align}
%
%The normalized steering vector from

Similar to the conventional fully passive IRS, it is assumed that each elements of hybrid IRS can independently reflect the incident signals. Let us denote the set of the $M_a$ active elements by $\Omega$. $\boldsymbol{\Theta}=\text{diag}\{\boldsymbol{\theta}^*\}=\text{diag}\{\theta_1, \cdots, \theta_m, \cdots, \theta_M\}\in \mathbb{C}^{M\times M}$, $\boldsymbol{\Psi}=\text{diag}\{\boldsymbol{\psi}^*\}\in \mathbb{C}^{M\times M}$, and $\boldsymbol{\Phi}=\text{diag}\{\boldsymbol{\phi}^*\}\in \mathbb{C}^{M\times M}$ represent the reflection coefficients of total elements, active elements, and passive elements of hybrid IRS, respectively, where
\begin{eqnarray}
\theta_m=
\left\{
\begin{array}{ll}
\ |\beta_m|e^{j\mu_m},\quad &\text{if}~ m\in \Omega,\\
\ e^{j\mu_m},\quad &\text{otherwise},
\end{array}
\right.
\end{eqnarray}
%\begin{eqnarray}
%\psi_m=
%\left\{
%\begin{array}{ll}
%\ |\beta_m|e^{j\mu_m},\quad &\text{if}~ m\in \Omega,\\
%\ 0,\quad &\text{otherwise},
%\end{array}
%\right.
%\end{eqnarray}
%and
%\begin{eqnarray}
%\phi_m=
%\left\{
%\begin{array}{ll}
%\ 0,\quad &\text{if}~ m\in \Omega,\\
%\ e^{j\mu_m},\quad &\text{otherwise},
%\end{array}
%\right.
%\end{eqnarray}
$\mu_m\in[0,2\pi)$ is the phase, and $|\beta_m|$ represents the amplifying coefficient, which is subject to the power of the IRS active elements.
Let us define
\begin{align}
\boldsymbol{\Psi}=\textbf{E}_{M_a}\boldsymbol{\Theta},~\boldsymbol{\Phi}={\textbf{E}}_{M_p}\boldsymbol{\Theta},
\end{align}
where
\begin{align}
\textbf{E}_{M_a}+{\textbf{E}}_{M_p}=\textbf{I}_M,~ \textbf{E}_{M_a}{\textbf{E}}_{M_p}=\textbf{0}_M,
\end{align}
the non-zero elements of the diagonal matrix $\textbf{E}_{M_a}\in\mathbb{C}^{M\times M}$ are unity whose positions are determined by $\Omega$.

The transmitted signal at BS is expressed as
\begin{align}
\textbf{s}=\sqrt{P}\textbf{v}x,
\end{align}  %N为总天线数
where $P$ represents the transmit power, $\textbf{v}\in\mathbb{C}^{N\times 1}$ and $x$ are the beamforming vector and the information symbol satisfying $\textbf{v}^H\textbf{v}=1$ and $\mathbb{E}[\|x\|^2]=1$, respectively.

In the presence of the PL, the received signal at  Bob is given by
\begin{align}\label{y_b}
y_b&=(\sqrt{\rho_{srb}}\textbf{h}^H_{rb}\boldsymbol{\Theta}\textbf{H}_{sr}+
\sqrt{\rho_{sb}}\textbf{h}^H_{sb})\textbf{s}+
\sqrt{\rho_{rb}}\textbf{h}^H_{rb}\boldsymbol{\Psi}\textbf{n}_r+n_b\nonumber\\
&=\sqrt{P}(\sqrt{\rho_{srb}}\textbf{h}^H_{rb}\boldsymbol{\Psi}\textbf{H}_{sr}+
\sqrt{\rho_{srb}}\textbf{h}^H_{rb}\boldsymbol{\Phi}\textbf{H}_{sr}+
\sqrt{\rho_{sb}}\textbf{h}^H_{sb})\textbf{v}x\nonumber\\
&~~~+\sqrt{\rho_{rb}}\textbf{h}^H_{rb}\boldsymbol{\Psi}\textbf{n}_r+n_b,
\end{align}
where $\rho_{srb}=\rho_{sr}\rho_{rb}$ represents the synthetic PL coefficient of BS-to-IRS channel and IRS-to-Bob channel, $\rho_{sb}$ and $\rho_{rb}$ stand for the PL coefficient of BS-to-Bob channel and IRS-to-Bob channel, respectively. $\textbf{n}_r\sim\mathcal {C}\mathcal {N}(\textbf{0}, \sigma^2_{r}\textbf{I}_{M_a})$ and $n_b\sim\mathcal {C}\mathcal {N}(0, \sigma^2_{b})$ denote the complex additive white Gaussian noise at the $M_a$ active elements of the hybrid IRS and at Bob, respectively.
$\textbf{h}_{sb}\in\mathbb{C}^{N\times 1}$, $\textbf{h}_{rb}\in\mathbb{C}^{M\times 1}$, and $\textbf{H}_{sr}=\textbf{h}_{rs}\textbf{h}^H_{sr}\in\mathbb{C}^{M\times N}$ represent the BS-to-Bob, IRS-to-Bob, and BS-to-IRS channels, respectively.
Let us define the channel $\textbf{h}_{tr}=\textbf{h}(\theta_{tr}, \varphi_{tr})$,
the normalized steering vector $\textbf{h}(\theta, \varphi)$ is expressed as
\begin{align}\label{h_theta}
&\textbf{h}(\theta, \varphi)\nonumber\\
&\buildrel \Delta \over=\frac{1}{\sqrt{N}}[e^{j2\pi\Phi_{\theta, \varphi}(1)}, \dots, e^{j2\pi\Phi_{\theta, \varphi}(n)}, \dots, e^{j2\pi\Phi_{\theta, \varphi}(N)}]^T,
\end{align}
where
\begin{align}
\Phi_{\theta, \varphi}(n)\buildrel \Delta \over =-\left(n-\frac{N+1}{2}\right)\frac{d \cos\theta\cos\varphi}{\lambda}, n=1, \dots, N,
\end{align}
$n$ stands for the antenna index, $d$ represents the spacing of adjacent transmitting antennas, $\theta$ means the direction angle of departure or arrival, $\varphi$ is the pitch angle, and $\lambda$ denotes the wavelength.

%\subsection{Problem Formulation}
In accordance with (\ref{y_b}), the achievable rate at Bob can be formulated as
\begin{align}\label{R_b}
&R_b=\log_2\left(1+\text{SNR}\right),
\end{align}
where
\begin{align}\label{SNR}
\text{SNR}=\frac{P|(\sqrt{\rho_{srb}}\textbf{h}^H_{rb}\boldsymbol{\Psi}\textbf{H}_{sr}+
\sqrt{\rho_{srb}}\textbf{h}^H_{rb}\boldsymbol{\Phi}\textbf{H}_{sr}+
\sqrt{\rho_{sb}}\textbf{h}^H_{sb})\textbf{v}|^2}
{\sigma_r^2|\sqrt{\rho_{rb}}\textbf{h}^H_{rb}\boldsymbol{\Psi}|^2+\sigma_b^2}.
\end{align}
The transmit power of all active elements at the hybrid IRS is given by
\begin{align}\label{pr0}
P_r=\text{Tr}\left(\boldsymbol{\Psi}\Big(\rho_{sr} P\textbf{H}_{sr}\textbf{v}\textbf{v}^H\textbf{H}^H_{sr}+
\sigma_r^2\textbf{I}_{M}\Big)\boldsymbol{\Psi}^H\right),
\end{align}
which satisfies $P_r\leq P^{\text{max}}_r$, where $P^{\text{max}}_r$ represents the maximum transmit power of $M_a$ active elements.

In this work, we maximize the SNR by jointly optimizing beamforming vector $\textbf{v}$, passive IRS PSM $\boldsymbol{\Phi}$, and active IRS PSM $\boldsymbol{\Psi}$. The overall optimization problem is formulated as follows
\vspace{-1.5mm}
\begin{subequations}\label{p0}
\begin{align}
&\max \limits_{\textbf{v}, \boldsymbol{\Phi}, \boldsymbol{\Psi}}
~~\text{SNR}\\
&~~\text{s.t.} ~~~\textbf{v}^H\textbf{v}=1, P_r\leq P^{\text{max}}_r, \label{P_r}\\
& ~~~~~~~~|\boldsymbol{\Phi}(m,m)|=1, \text{if}~ m\not\in \Omega,\\
&~~~~~~~~|\boldsymbol{\Phi}(m,m)|=0, \text{otherwise},\label{phi5}\\
& ~~~~~~~~|\boldsymbol{\Psi}(m,m)|\leq \beta_{\text{max}}, \text{if}~m\in \Omega, \label{psi51}\\
&~~~~~~~~|\boldsymbol{\Psi}(m,m)|=0, \text{otherwise}, \label{psi5}
\end{align}
\end{subequations}
where $\beta_{\text{max}}$ is the amplitude budget. Considering that this optimization problem is a non-convex problem with a constant modulus constraint, and it is challenging to tackle it directly in general. In what follows, the alternating optimization algorithm is proposed to compute the beamforming vector and hybrid IRS PSM, respectively.

\section{Proposed Max-SNR-FP scheme}\label{s2}
In this section, we construct a Max-SNR-FP algorithm to jointly optimize the beamforming vector $\textbf{v}$, passive IRS PSM $\boldsymbol{\Phi}$, and active IRS PSM $\boldsymbol{\Psi}$. In what follows, we will alternately solve for $\textbf{v}$, $\boldsymbol{\Phi}$, and $\boldsymbol{\Psi}$.

\subsection{Optimize $\textbf{v}$ given $\boldsymbol{\Phi}$ and $\boldsymbol{\Psi}$}
Firstly, we transform the power constraint in (\ref{P_r}) into a convex constraint with respect to $\textbf{v}$ as follows
\begin{align}\label{p_r_v}
P_r=\textbf{v}^H\left(\rho_{sr} P \textbf{H}^H_{sr}\boldsymbol{\Psi}^H\boldsymbol{\Psi}\textbf{H}_{sr}\right)\textbf{v}+
\text{Tr}\left(\sigma_r^2\boldsymbol{\Psi}\boldsymbol{\Psi}^H\right)\leq P^{\text{max}}_r.
\end{align}
Then, given $\boldsymbol{\Phi}$ and $\boldsymbol{\Psi}$, the optimal beamforming vector $\textbf{v}$ can be found by addressing the problem in what follows
\begin{subequations}\label{w_u}
\begin{align}
&~\max \limits_{\textbf{v}}
~\textbf{v}^H\textbf{A}\bar{\textbf{v}}\\
&~~~\text{s.t.}~~\textbf{v}^H\textbf{v}=1,(\ref{p_r_v}),
\end{align}
\end{subequations}
where
\begin{align}
\textbf{A}=&(\sqrt{\rho_{srb}}\textbf{h}^H_{rb}\boldsymbol{\Phi}\textbf{H}_{sr}+
\sqrt{\rho_{srb}}\textbf{h}^H_{rb}\boldsymbol{\Psi}\textbf{H}_{sr}+\sqrt{\rho_{sb}}\textbf{h}^H_{sb})^H\nonumber\\
&(\sqrt{\rho_{srb}}\textbf{h}^H_{rb}\boldsymbol{\Phi}\textbf{H}_{sr}+
\sqrt{\rho_{srb}}\textbf{h}^H_{rb}\boldsymbol{\Psi}\textbf{H}_{sr}+\sqrt{\rho_{sb}}\textbf{h}^H_{sb}).
\end{align}
It is clear that this problem is not convex, and in accordance with the Taylor series expansion, we have
\begin{align}
\textbf{v}^H\textbf{A}\textbf{v}\geq 2\Re\{\bar{\textbf{v}}^H\textbf{A}\textbf{v}\}-\bar{\textbf{v}}^H\textbf{A}\bar{\textbf{v}},
\end{align}
where $\bar{\textbf{v}}$ is a given vector. Then (\ref{w_u}) can be recasted as
\begin{subequations}\label{w}
\begin{align}
&~\max \limits_{\textbf{v}}
~2\Re\{\bar{\textbf{v}}^H\textbf{A}\textbf{v}\}-\bar{\textbf{v}}^H\textbf{A}\bar{\textbf{v}} \\ &~~~\text{s.t.}~~\textbf{v}^H\textbf{v}=1,(\ref{p_r_v}).
\end{align}
\end{subequations}
Given that this optimization problem is convex, we can obtain the optimal $\textbf{v}$ by adopting the CVX tool.
\subsection{Optimize $\boldsymbol{\Phi}$ given $\textbf{v}$ and  $\boldsymbol{\Psi}$}
In order to simplify the SNR expression with respect to the PSM $\boldsymbol{\Phi}$, we treat $\textbf{v}$ and  $\boldsymbol{\Psi}$ as two constants, and define
\begin{align}
B=(\sqrt{\rho_{srb}}\textbf{h}^H_{rb}\boldsymbol{\Psi}\textbf{H}_{sr}+
\sqrt{\rho_{sb}}\textbf{h}^H_{sb})\textbf{v}.
\end{align}
Then, the subproblem to optimize PSM $\boldsymbol{\Phi}$ is
\begin{subequations}\label{Phi_1}
\begin{align}
&~\max \limits_{\boldsymbol{\Phi}}
~|\sqrt{\rho_{srb}}\textbf{h}^H_{rb}\boldsymbol{\Phi}\textbf{H}_{sr}\textbf{v}+B|^2\\
&~~~\text{s.t.}~~|\boldsymbol{\Phi}(m,m)|=1,\text{if}~ m\not\in \Omega, \label{relax}\\
&~~~~~~~~|\boldsymbol{\Phi}(m,m)|=0, \text{otherwise}. \label{model0}
\end{align}
\end{subequations}
By defining
\begin{align}
\textbf{C}=\rho_{srb}\text{diag}\{\textbf{h}^H_{rb}\}\textbf{H}_{sr}\textbf{v}
\textbf{v}^H\textbf{H}^H_{sr}\text{diag}\{\textbf{h}^H_{rb}\}^H,
\end{align}
and based on the fact that diag$\{\textbf{p}\}\textbf{q}=\text{diag}\{\textbf{q}\}\textbf{p}$ for $\textbf{p}, \textbf{q}\in \mathbb{C}^{M\times 1}$, the objective function in (\ref{Phi_1}) can be recasted as
\begin{align}
\boldsymbol{\phi}^H\textbf{C}\boldsymbol{\phi}+
2\Re\{\sqrt{\rho_{srb}}\boldsymbol{\phi}^H\text{diag}\{\textbf{h}^H_{rb}\}\textbf{H}_{sr}\textbf{v}B^*\}+|B|^2.
\end{align}
According to the Taylor series expansion, we have
\begin{align}
\boldsymbol{\phi}^H\textbf{C}\boldsymbol{\phi}\geq 2\Re\{\bar{\boldsymbol{\phi}}^H\textbf{C}\boldsymbol{\phi}\}-\bar{\boldsymbol{\phi}}^H\textbf{C}\bar{\boldsymbol{\phi}},
\end{align}
where $\bar{\boldsymbol{\phi}}$ is a given vector. In addition, similar to the results in \cite{Shi2021Secure}, the unit modulus constraint (\ref{relax}) can be relaxed to
\begin{align}\label{relax1}
|\boldsymbol{\Phi}(m,m)|\leq1, \text{if}~ m\not\in \Omega.
\end{align}
At this point, the subproblem (\ref{Phi_1}) can be rewritten as follows
\begin{subequations}\label{Phi_2}
\begin{align}
&~\max \limits_{\boldsymbol{\Phi}}
~2\Re\{\bar{\boldsymbol{\phi}}^H\textbf{C}\boldsymbol{\phi}\}-\bar{\boldsymbol{\phi}}^H\textbf{C}\bar{\boldsymbol{\phi}}
+|B|^2+2\Re\{\sqrt{\rho_{srb}}\boldsymbol{\phi}^H \bullet\nonumber\\
&~~~~~~~~~\text{diag}\{\textbf{h}^H_{rb}\}\textbf{H}_{sr}\textbf{v}B^*\}\\
&~~~\text{s.t.}~~~ (\ref{relax1}), (\ref{model0}).
\end{align}
\end{subequations}
This problem can be solved directly with the convex optimization tool since it is convex.

\subsection{Optimize $\boldsymbol{\Psi}$ given $\textbf{v}$ and  $\boldsymbol{\Phi}$}
To optimize $\boldsymbol{\Psi}$, we regard $\textbf{v}$ and  $\boldsymbol{\Phi}$ as two given constants, and transform  the power constraint in (\ref{P_r}) into a convex constraint on $\boldsymbol{\psi}$ as follows
\begin{align}\label{p_max}
P_r&=\text{Tr}\left(\boldsymbol{\Psi}\Big(\rho_{sr} P\textbf{H}_{sr}\textbf{v}\textbf{v}^H\textbf{H}^H_{sr}+
\sigma_r^2\textbf{I}_{M}\Big)\boldsymbol{\Psi}^H\right)\nonumber\\
&=\boldsymbol{\psi}^T(\rho_{sr} P \text{diag}\{\textbf{v}^H\textbf{H}^H_{sr}\}
\text{diag}\{\textbf{H}_{sr}\textbf{v}\}+\sigma_r^2\textbf{I}_{M})\boldsymbol{\psi}^* \nonumber\\
&\leq P^{\text{max}}_r .
\end{align}
By neglecting the constant terms, the subproblem with respect to $\boldsymbol{\Psi}$ is given by
\begin{subequations}\label{Psi_1}
\begin{align}
&~\max \limits_{\boldsymbol{\Psi}}
~\frac{|(\sqrt{\rho_{srb}}\textbf{h}^H_{rb}\boldsymbol{\Psi}\textbf{H}_{sr}+
\sqrt{\rho_{srb}}\textbf{h}^H_{rb}\boldsymbol{\Phi}\textbf{H}_{sr}+
\sqrt{\rho_{sb}}\textbf{h}^H_{sb})\textbf{v}|^2}
{\sigma_r^2|\sqrt{\rho_{rb}}\textbf{h}^H_{rb}\boldsymbol{\Psi}|^2+\sigma_b^2}\\
&~~~\text{s.t.}~~~(\ref{psi51}), (\ref{psi5}), (\ref{p_max}).
\end{align}
\end{subequations}
Let us define
\begin{align}
D=(\sqrt{\rho_{srb}}\textbf{h}^H_{rb}\boldsymbol{\Phi}\textbf{H}_{sr}+
\sqrt{\rho_{sb}}\textbf{h}^H_{sb})\textbf{v}.
\end{align}
Then, the objective function in (\ref{Psi_1}) can be converted to
\begin{align}\label{Psi_3}
\frac{\boldsymbol{\psi}^H\textbf{C}\boldsymbol{\psi}+
2\Re\{\boldsymbol{\psi}^H\sqrt{\rho_{srb}}\text{diag}\{\textbf{h}^H_{rb}\}\textbf{H}_{sr}\textbf{v}D^*\}+|D|^2}
{\sigma_r^2\rho_{rb}|\boldsymbol{\psi}^H\text{diag}\{\textbf{h}^H_{rb}\}|^2+\sigma_b^2}.
\end{align}
At this point, the optimization problem (\ref{Psi_1}) has become a nonlinear fractional optimization problem. Based on  the FP strategy in \cite{Dinkelbach1967On}, we introduce a parameter $\tau$ and transform the objective function (\ref{Psi_3}) as
\begin{align}
&\boldsymbol{\psi}^H\textbf{C}\boldsymbol{\psi}+
2\Re\{\boldsymbol{\psi}^H\sqrt{\rho_{srb}}\text{diag}\{\textbf{h}^H_{rb}\}\textbf{H}_{sr}\textbf{v}D^*\}+|D|^2
\nonumber\\
&-\tau(\sigma_r^2\rho_{rb}|\boldsymbol{\psi}^H\text{diag}\{\textbf{h}^H_{rb}\}|^2+\sigma_b^2).
\end{align}
The optimal solution can be achieved if and only if $\boldsymbol{\psi}^H\textbf{C}\boldsymbol{\psi}+
2\Re\{\boldsymbol{\psi}^H\sqrt{\rho_{srb}}\text{diag}\{\textbf{h}^H_{rb}\}\textbf{H}_{sr}\textbf{v}D^*\}+|D|^2
-\tau(\sigma_r^2\rho_{rb}|\boldsymbol{\psi}^H\text{diag}\{\textbf{h}^H_{rb}\}|^2+\sigma_b^2)=0$.
We linearize the $\boldsymbol{\psi}^H\textbf{C}\boldsymbol{\psi}$ by employing Taylor series expansion at a given vector $\bar{\boldsymbol{\psi}}$, the subproblem with respect to $\boldsymbol{\Psi}$ can be recasted as
%\begin{subequations}
\begin{align}\label{Psi_2}
&~\max \limits_{\boldsymbol{\Psi}, \tau}
~~2\Re\{\bar{\boldsymbol{\psi}}^H\textbf{C}\boldsymbol{\psi}\}-\bar{\boldsymbol{\psi}}^H\textbf{C}\bar{\boldsymbol{\psi}}+
2\Re\{\boldsymbol{\psi}^H\sqrt{\rho_{srb}}\text{diag}\{\textbf{h}^H_{rb}\}\bullet
\nonumber\\
&~~~~~~~~~\textbf{H}_{sr}\textbf{v}D^*\}+|D|^2-
\tau(\sigma_r^2\rho_{rb}|\boldsymbol{\psi}^H\text{diag}\{\textbf{h}^H_{rb}\}|^2+\sigma_b^2)
\nonumber\\
&~~~\text{s.t.}~~~(\ref{psi51}), (\ref{psi5}), (\ref{p_max}).
\end{align}
%\end{subequations}
Notice that problem (\ref{Psi_2}) is convex, and it can be addressed effectively with the CVX tool.
%\subsection{Overall Algorithm}
%So far, we have completed the design of $\textbf{v}$, $\boldsymbol{\Phi}$, and $\boldsymbol{\Psi}$. The iterative idea of the proposed Max-SNR-FP scheme is summarized as follows: given RIS phase shift matrices $\boldsymbol{\Phi}$ and $\boldsymbol{\Psi}$, the beamforming $\textbf{v}$ can be computed by using Taylor series expansion; given $\textbf{v}$ and $\boldsymbol{\Psi}$, $\boldsymbol{\Phi}$ is solved by employing Taylor series expansion; given $\textbf{v}$ and $\boldsymbol{\Phi}$, $\boldsymbol{\Psi}$ can be obtained by FP method. The alternative iteration process among $\textbf{v}$, $\boldsymbol{\Phi}$, and $\boldsymbol{\Psi}$ is repeated until the termination condition is met, i.e., $|R_b^{(p)}-R_b^{(p-1)}\big|\leq\epsilon$, where $p$ and $\epsilon$ are the iteration index and threshold, respectively.
The whole procedure of the Max-SNR-FP scheme is described in Algorithm 1.
\begin{algorithm}
\caption{Proposed Max-SNR-FP algorithm}
\begin{algorithmic}[1]
\STATE Initialize feasible solutions $\textbf{v}^{(0)}$, $\boldsymbol{\Phi}^{(0)}$, and $\boldsymbol{\Psi}^{(0)}$, calculate achievable rate $R^{(0)}_b$ based on (\ref{R_b}).
\STATE Set the iteration number $k=0$, accuracy value $\epsilon$.
\REPEAT
\STATE Given $\boldsymbol{\Phi}^{(k)}$ and $\boldsymbol{\Psi}^{(k)}$, solve (\ref{w}) to get $\textbf{v}^{(k+1)}$.
\STATE Given $\textbf{v}^{(k+1)}$ and $\boldsymbol{\Psi}^{(k)}$, solve (\ref{Phi_2}) to get $\boldsymbol{\Phi}^{(k+1)}$.
\STATE Given $\textbf{v}^{(k+1)}$ and $\boldsymbol{\Phi}^{(k+1)}$, solve (\ref{Psi_2}) to get $\boldsymbol{\Psi}^{(k+1)}$.
\STATE Calculate $R^{(k+1)}_b$ based on $\textbf{v}^{(k+1)}$, $\boldsymbol{\Phi}^{(k+1)}$, and $\boldsymbol{\Psi}^{(k+1)}$.
\STATE Update $k=k+1$.
\UNTIL {$|R_b^{(k)}-R_b^{(k-1)}|\leq\epsilon$.}
\end{algorithmic}
\end{algorithm}

The overall computational complexity of the proposed Max-SNR-FP algorithm is
$\mathcal {O}(L((M+1)^3+2MN^2+2M^2)\text{In}(1/\epsilon)+M^3+N^3+5M^2+2MN+2M+2MN^2)$
float-point operations (FLOPs), where $L$ represents the numbers of alternating iterations, $\epsilon$ denotes the accuracy.
%%

%%

%\textcolor{red}{
%\begin{align}\label{}
%&R_b(\textbf{v}, \boldsymbol{\Theta})
%=\log_2\left(\frac{\alpha P\text{Tr}(\widehat{\boldsymbol{\Theta}}\widehat{\textbf{H}}_{BB})+
%\sigma^2\rho_{rb}\text{Tr}(\widehat{\boldsymbol{\Theta}}\widehat{\textbf{H}}_{RB})+\sigma^2}
%{\sigma^2\rho_{rb}\text{Tr}(\widehat{\boldsymbol{\Theta}}\widehat{\textbf{H}}_{RB})+\sigma^2}\right)\nonumber\\
%&=\log_2\left(\frac{\alpha P(\text{vec}(\widehat{\textbf{H}}^T_{BB}))^T{\boldsymbol{t}}+
%\sigma^2\rho_{rb}(\text{vec}(\widehat{\textbf{H}}^T_{RB}))^T{\boldsymbol{t}}+\sigma^2}
%{\sigma^2\rho_{rb}(\text{vec}(\widehat{\textbf{H}}^T_{RB}))^T{\boldsymbol{t}}+\sigma^2}\right),
%\end{align}
%\begin{align}\label{}
%&R_e(\textbf{v}, \boldsymbol{\Theta})=\nonumber\\
%&=\log_2\Big(\nonumber\\
%&\frac{\alpha P(\text{vec} (\widehat{\textbf{H}}^T_{EE}))^T{\boldsymbol{t}}
%+\sigma^2\rho_{re}(\text{vec}(\widehat{\textbf{H}}^T_{RE}))^T{\boldsymbol{t}}+(1-\alpha)\rho_{se} P|\textbf{h}^H_{se}\textbf{T}_{AN}|^2+\sigma^2}
%{\sigma^2\rho_{re}(\text{vec}(\widehat{\textbf{H}}^T_{RE}))^T{\boldsymbol{t}}+(1-\alpha)\rho_{se} P|\textbf{h}^H_{se}\textbf{T}_{AN}|^2+\sigma^2}\Big),
%\end{align}
%where $\boldsymbol{t}=\text{vec}(\widehat{\boldsymbol{\Theta}})$.}

\section{Proposed Max-SNR-EAR scheme}\label{s3}
In the previous section, we proposed the Max-SNR-FP method to compute the beamforming vector $\textbf{v}$, IRS phase shift matrices $\boldsymbol{\Phi}$ and $\boldsymbol{\Psi}$. However, it comes with a high computational complexity. To decrease the complexity, the Max-SNR-EAR scheme with low-complexity is proposed in this section.

\subsection{Optimize $\textbf{v}$ given $\boldsymbol{\Phi}$ and $\boldsymbol{\Psi}$}
Given IRS phase shift matrices $\boldsymbol{\Phi}$ and $\boldsymbol{\Psi}$, in accordance with the principle of maximizing SLNR in \cite{Sadek2007A}, the beamforming vector $\textbf{v}$ can be optimized by tackling the problem in what follows
\begin{subequations}\label{v2}
\begin{align}
&~\max \limits_{\textbf{v}}
~\text{SLNR}=\frac{\textbf{v}^H\textbf{E}\textbf{v}}{\textbf{v}^H(\sigma_b^2\textbf{I}_N)\textbf{v}}\\
&~~~\text{s.t.}~~\textbf{v}^H\textbf{v}=1,(\ref{p_r_v}),
\end{align}
\end{subequations}
where
\begin{align}
\textbf{E}=&\rho_{srb}\textbf{H}_{sr}^H\boldsymbol{\Phi}^H\textbf{h}_{rb}\textbf{h}^H_{rb}\boldsymbol{\Phi}\textbf{H}_{sr}+
\rho_{srb}\textbf{H}_{sr}^H\boldsymbol{\Psi}^H\textbf{h}_{rb}\textbf{h}^H_{rb}\boldsymbol{\Psi}\textbf{H}_{sr}\nonumber\\
&+\textbf{h}_{sb}\textbf{h}^H_{sb}.
\end{align}
According to the Taylor series expansion and neglecting the constant terms, the problem (\ref{v2}) can be recasted as
\begin{subequations}\label{w2}
\begin{align}
&~\max \limits_{\textbf{v}}
~2\Re\{\bar{\textbf{v}}^H\textbf{E}\textbf{v}\}-\bar{\textbf{v}}^H\textbf{E}\bar{\textbf{v}}\\
&~~~\text{s.t.}~~~\textbf{v}^H\textbf{v}=1,(\ref{p_r_v}),
\end{align}
\end{subequations}
which can be addressed directly via adopting the convex optimization tool.
\subsection{Optimize $\boldsymbol{\Phi}$ and  $\boldsymbol{\Psi}$ given $\textbf{v}$}
Given beamforming vector $\textbf{v}$, we consider to design the phase of hybrid IRS firstly. The confidential message received by Bob through the cascade path is expressed as
\begin{align}
P\rho_{srb}\textbf{h}_{rb}^H\boldsymbol{\Theta}\textbf{H}_{sr}\textbf{v}
\textbf{v}^H\textbf{H}^H_{sr}\boldsymbol{\Theta}^H\textbf{h}_{rb}.
\end{align}
To maximize the confidential message of the cascade path, the phase alignment strategy is employed to design the hybrid IRS phase $\widetilde{\boldsymbol{\theta}}$,  $\widetilde{\boldsymbol{\theta}}$ is given by
\begin{align}\label{w_theta}
\widetilde{\boldsymbol{\theta}}=[e^{(-i\text{arg}(\textbf{s}_1))}, \cdots, e^{(-i\text{arg}(\textbf{s}_M))}]^T,
\end{align}
where $\textbf{s}=\text{diag}\{\textbf{h}_{rb}^H\}\textbf{H}_{sr}\textbf{v}$, and $\textbf{s}_k$ is the $k$-th element of $\textbf{s}$.

Next, inspired by the amplitude design of fully active IRS in \cite{Zhang2021Active}, we suppose that all active elements of the hybrid IRS have the same amplitude. Based on the IRS power constraint in (\ref{P_r}), we have
\begin{align}\label{Q}
|\beta|=\sqrt{{P^{\max}_r}/{Q}},
\end{align}
where
\begin{align}
Q=&\text{Tr}(\widetilde{\boldsymbol{\theta}}^H(\rho_{sr}P\text{diag}\{\textbf{v}^H\textbf{H}^H_{sr}\textbf{E}_{M_a}\}
\text{diag}\{\textbf{v}^H\textbf{H}^H_{sr}\textbf{E}_{M_a}\}^H \nonumber\\
&+\sigma^2\textbf{E}_{M_a}\textbf{E}_{M_a})\widetilde{\boldsymbol{\theta}}).
\end{align}
Based on (\ref{w_theta}) and (\ref{Q}), we obtain the passive IRS PSM and active IRS PSM as follows
\begin{align}
&\boldsymbol{\Phi}=\textbf{E}_{M_p}\text{diag}\{\widetilde{\boldsymbol{\theta}}\},\\
&\boldsymbol{\Psi}=|\beta|\textbf{E}_{M_a}\text{diag}\{\widetilde{\boldsymbol{\theta}}\}.
\end{align}
%\subsection{Overall Algorithm}
%Now, we have complected the design of beamforming vector $\textbf{v}$ and RIS phase shift matrices $\boldsymbol{\Phi}$ and  $\boldsymbol{\Psi}$. The key idea of proposed Max-SNR-EAR algorithm is summarized as follows: first, fix $\boldsymbol{\Phi}$ and  $\boldsymbol{\Psi}$, and use the principle of maximizing SLNR to obtain $\textbf{v}$; next, given $\textbf{v}$, and employ the idea of maximum ratio reflecting and EAR to construct $\boldsymbol{\Phi}$ and  $\boldsymbol{\Psi}$; finally, loop the above steps, and calculate $\textbf{v}$, $\boldsymbol{\Phi}$, and  $\boldsymbol{\Psi}$ alternately until convergence.

Similar to Algorithm 1, we calculate $\textbf{v}$, $\boldsymbol{\Phi}$, and  $\boldsymbol{\Psi}$ alternately until convergence, i.e., $|R_b^{(k)}-R_b^{(k-1)}|\leq\epsilon$.
The overall computational complexity of Max-SNR-EAR scheme is
%\begin{align}
%&\mathcal {O}(K((M+1)^3+2MN^2+2M^2)\text{In}(1/k_1)+((M+1)^3\nonumber\\
%&~~~+M^2)\text{In}(1/k_2)+N^3+2MN+2M),
%\end{align}
$\mathcal {O}(K(2M^2+N^3+8N^2M+2MN)$
FLOPs, where $K$ is the numbers of alternating iterations.

%\begin{algorithm}
%\caption{Proposed Max-SNR-LK algorithm}
%\begin{algorithmic}[1]
%\STATE Initialize $\textbf{v}^{(0)}$, $\boldsymbol{\Phi}^{(0)}$, and $\boldsymbol{\Psi}^{(0)}$, compute $R^{(0)}_b$ according to (\ref{R_b}).
%\STATE Set $p=0$, threshold value $\epsilon$.
%\REPEAT
%\STATE Given $\boldsymbol{\Phi}^{(p)}$ and $\boldsymbol{\Psi}^{(p)}$, solve problem (\ref{w2}) to determine $\textbf{v}^{(p+1)}$.
%\STATE Given $\textbf{v}^{(p+1)}$ and $\boldsymbol{\Psi}^{(p)}$, solve problem (\ref{Phi_4}) to determine $\boldsymbol{\Phi}^{(p+1)}$.
%\STATE Given $\textbf{v}^{(p+1)}$ and $\boldsymbol{\Phi}^{(p)}$, solve problem (\ref{Psi_2}) to determine $\boldsymbol{\Psi}^{(p+1)}$.
%\STATE Compute $R^{(p+1)}_b$ using $\textbf{v}^{(p+1)}$, $\boldsymbol{\Phi}^{(p+1)}$, and $\boldsymbol{\Psi}^{(p+1)}$.
%\STATE $p=p+1$.
%\UNTIL {$\big|R_b^{(p)}-R_b^{(p-1)}\big|\leq\epsilon$.}
%\end{algorithmic}
%\end{algorithm}

\section{Proposed Max-SNR-MM scheme}\label{s4}

In the Section \ref{s2} and Section \ref{s3}, the Max-SNR-FP and Max-SNR-EAR schemes have been presented to jointly calculate the transmit beamforming vector and IRS phase shift matrices, where the active and passive IRS phase shift matrices are optimized respectively. To investigate the effect of joint design of active and passive IRS phase shift matrices on the system performance enhancement, in this section,  we propose a low-complexity scheme, named Max-SNR-MM, to boost the achievable rate. In the following, based on the criteria of alternate optimization, we will alternately solve for $\textbf{v}$ and $\boldsymbol{\Theta}$.

\subsection{Optimize $\textbf{v}$ given $\boldsymbol{\Theta}$}
In accordance with (\ref{y_b}), the received signal can be rewritten as
\begin{align}\label{}
y_b&=(\sqrt{\rho_{srb}}\textbf{h}^H_{rb}\boldsymbol{\Theta}\textbf{H}_{sr}+
\sqrt{\rho_{sb}}\textbf{h}^H_{sb})\textbf{s}+
\sqrt{\rho_{rb}}\textbf{h}^H_{rb}\boldsymbol{\Psi}\textbf{n}_r+n_b\nonumber\\
&=\sqrt{P}(\sqrt{\rho_{srb}}\textbf{h}^H_{rb}\boldsymbol{\Theta}\textbf{H}_{sr}+
\sqrt{\rho_{sb}}\textbf{h}^H_{sb})\textbf{v}x+\sqrt{\rho_{rb}}\textbf{h}^H_{rb}
\textbf{E}_{M_a}\boldsymbol{\Theta}\textbf{n}_r\nonumber\\
&~~~+n_b.
\end{align}
Then, the SNR can be expressed as
\begin{align}\label{}
\text{SNR}=\frac{P|(\sqrt{\rho_{srb}}\textbf{h}^H_{rb}\boldsymbol{\Theta}\textbf{H}_{sr}+
\sqrt{\rho_{sb}}\textbf{h}^H_{sb})\textbf{v}|^2}
{\sigma_r^2|\sqrt{\rho_{rb}}\textbf{h}^H_{rb}\textbf{E}_{M_a}\boldsymbol{\Theta}|^2+\sigma_b^2}.
\end{align}

The transmit power of the hybrid IRS in (\ref{P_r}) can be reformulated as
\begin{align}\label{P_r2}
P_r&=\text{Tr}\left(\textbf{E}_{M_a}\boldsymbol{\Theta}\Big(\rho_{sr} P\textbf{H}_{sr}\textbf{v}\textbf{v}^H\textbf{H}^H_{sr}+
\sigma_r^2\textbf{I}_{M}\Big)\textbf{E}_{M_a}\boldsymbol{\Theta}^H\right)\nonumber\\
&=\textbf{v}^H(\rho_{sr}P\textbf{H}^H_{sr}\textbf{E}_{M_a}\boldsymbol{\Theta}^H
\textbf{E}_{M_a}\boldsymbol{\Theta}\textbf{H}^H_{sr})\textbf{v}+
\text{Tr}(\sigma_r^2\textbf{E}_{M_a}\boldsymbol{\Theta}\bullet\nonumber\\
&~~~\textbf{E}_{M_a}\boldsymbol{\Theta}^H)\nonumber\\
&\leq P^{\text{max}}_r.
\end{align}
Given IRS phase shift matrix $\boldsymbol{\Theta}$, the optimization problem with respect to \textbf{v} is given by
\begin{subequations}\label{v31}
\begin{align}
&\max \limits_{\textbf{v}}
~~\frac{P|(\sqrt{\rho_{srb}}\textbf{h}^H_{rb}\boldsymbol{\Theta}\textbf{H}_{sr}+
\sqrt{\rho_{sb}}\textbf{h}^H_{sb})\textbf{v}|^2}
{\sigma_r^2|\sqrt{\rho_{rb}}\textbf{h}^H_{rb}\textbf{E}_{M_a}\boldsymbol{\Theta}|^2+\sigma_b^2}\\
&~~\text{s.t.} ~~~\textbf{v}^H\textbf{v}=1, (\ref{P_r2}).
\end{align}
\end{subequations}
It is observed that the form of the problem (\ref{v31}) is similar to that of the problem (\ref{w_u}), and the same method is taken into account to solve for problem (\ref{v31}). For the sake of brevity, it is not repeated here.

\subsection{Optimize $\boldsymbol{\Theta}$  given $\textbf{v}$}
Given beamforming vector $\textbf{v}$, we focus on optimizing the hybrid IRS phase shift matrix $\boldsymbol{\Theta}$ in this subsection.
Let us define
\begin{align}
&\widehat{\boldsymbol{\theta}}=[\boldsymbol{\theta}; 1], \\
&\textbf{h}_b=\left[ \begin{array}{*{20}{c}}
\sqrt{\rho_{srb}P}\text{diag}\{\textbf{h}^H_{rb}\}\textbf{H}_{sr}\textbf{v}\\
\sqrt{\rho_{sb}P}\textbf{h}^H_{sb}\textbf{v}\end{array}\right]_{(M+1)\times 1}, \\
&\textbf{H}_e=\left[ \begin{array}{*{20}{c}}
\sqrt{\rho_{rb}}\text{diag}\{\textbf{h}^H_{rb}\textbf{E}_{M_a}\}\\
\textbf{0}_{1\times M}
\end{array}\right]_{(M+1)\times M}.
\end{align}
Then, the SNR in (\ref{SNR}) can be reformulated as follows
\begin{align}
\text{SNR}&=\frac{P|(\sqrt{\rho_{srb}}\textbf{h}^H_{rb}\boldsymbol{\Psi}\textbf{H}_{sr}+
\sqrt{\rho_{srb}}\textbf{h}^H_{rb}\boldsymbol{\Phi}\textbf{H}_{sr}+
\sqrt{\rho_{sb}}\textbf{h}^H_{sb})\textbf{v}|^2}
{\sigma_r^2|\sqrt{\rho_{rb}}\textbf{h}^H_{rb}\boldsymbol{\Psi}|^2+\sigma_b^2}\nonumber\\
&=\frac{P|(\sqrt{\rho_{srb}}\textbf{h}^H_{rb}\boldsymbol{\Theta}\textbf{H}_{sr}+
\sqrt{\rho_{sb}}\textbf{h}^H_{sb})\textbf{v}|^2}
{\sigma_r^2|\sqrt{\rho_{rb}}\textbf{h}^H_{rb}\textbf{E}_{M_a}\boldsymbol{\Theta}|^2+\sigma_b^2}\nonumber\\
&=\frac{|\widehat{\boldsymbol{\theta}}^H\textbf{h}_b|^2}{\sigma^2_r|\widehat{\boldsymbol{\theta}}^H\textbf{H}_e|^2+\sigma_b^2}.
\end{align}
Accordingly, the power constraint in (\ref{p_max}) can be rewritten as
\begin{align}\label{Pr2}
P_r&=\text{Tr}\left(\boldsymbol{\Psi}\Big(\rho_{sr} P\textbf{H}_{sr}\textbf{v}\textbf{v}^H\textbf{H}^H_{sr}+
\sigma_r^2\textbf{I}_{M}\Big)\boldsymbol{\Psi}^H\right)\nonumber\\
&= \text{Tr}\left(\textbf{E}_{M_a}\boldsymbol{\Theta}\Big(\rho_{sr} P\textbf{H}_{sr}\textbf{v}\textbf{v}^H\textbf{H}^H_{sr}+
\sigma_r^2\textbf{I}_{M}\Big)\textbf{E}_{M_a}\boldsymbol{\Theta}^H\right)\nonumber\\
&=\boldsymbol{\theta}^{T}(\rho_{sr}P\textbf{E}_{M_a}\text{diag}\{\textbf{v}^H\textbf{H}^H_{sr}\}
\text{diag}\{\textbf{H}_{sr}\textbf{v}\}\textbf{E}_{M_a}+ \nonumber\\
&~~~\sigma_r^2\textbf{E}_{M_a}\textbf{E}_{M_a})\boldsymbol{\theta}^*\nonumber\\
&=\widehat{\boldsymbol{\theta}}^T\text{blkdiag}\{\rho_{sr}P\textbf{E}_{M_a}\text{diag}\{\textbf{v}^H\textbf{H}^H_{sr}\}
\text{diag}\{\textbf{H}_{sr}\textbf{v}\}\textbf{E}_{M_a}+\nonumber\\
&~~~\sigma_r^2\textbf{E}_{M_a}\textbf{E}_{M_a};0 \}\widehat{\boldsymbol{\theta}}^*\nonumber\\
&\leq P^{\text{max}}_r .
\end{align}
Then, the optimization problem (\ref{p0}) regarding $\boldsymbol{\Theta}$ can be recasted as follows
\begin{subequations}\label{SNR3}
\begin{align}
&\max \limits_{\widehat{\boldsymbol{\theta}}}
~~\frac{|\widehat{\boldsymbol{\theta}}^H\textbf{h}_b|^2}{\sigma^2_r|\hat{\boldsymbol{\theta}}^H\textbf{H}_e|^2+\sigma_b^2}\\
&~~\text{s.t.} ~~~|\widehat{\boldsymbol{\theta}}(m)|=1, \text{if}~ m\not\in \Omega,\\
& ~~~~~~~~|\widehat{\boldsymbol{\theta}}(m)|\leq \beta_{\text{max}}, \text{if}~m\in \Omega, \\
&~~~~~~~~|\widehat{\boldsymbol{\theta}}(m+1)|=1, ~(\ref{Pr2}).
\end{align}
\end{subequations}
Based on the result in \cite{Nasir2017Secrecy}, i.e.,
\begin{align}
&\text{In}\left(1+\frac{|\alpha|^2}{\gamma}\right)\geq \text{In}\left(1+\frac{|\bar{\alpha}|^2}{\bar{\gamma}}\right)-\frac{|\bar{\alpha}|^2}{\bar{\gamma}}+\nonumber\\
&~~~~~~~~~~~~~~~~~~~~~\frac{2\Re\{\bar{\alpha}\alpha\}}{\bar{\gamma}}-\frac{|\bar{\alpha}|^2(\gamma+|\alpha|^2)}
{\bar{\gamma}(\bar{\gamma}+|\bar{\alpha}|^2)},
\end{align}
where $\bar{\alpha}$ and $\bar{\gamma}$ represent the fixed points, by neglecting the constant terms, the objective function in (\ref{SNR3}) can be reformulated as
\begin{align}
\frac{2\Re\{\widehat{\boldsymbol{\theta}}^H\textbf{h}_b\textbf{h}^H_b\widehat{\boldsymbol{\theta}}\}}{c}-
\frac{d(\sigma_r^2|\widehat{\boldsymbol{\theta}}^H\textbf{H}_e|^2+\sigma_b^2+|\widehat{\boldsymbol{\theta}}^H\textbf{h}_b|^2)}{c(c+d)},
\end{align}
where $c=\sigma_r^2|\bar{\boldsymbol{\theta}}^H\textbf{H}_e|^2+\sigma_b^2$, $d=|\bar{\boldsymbol{\theta}}^H\textbf{h}_b|^2$, and $\bar{\boldsymbol{\theta}}$ stands for the fixed point. Let us define
\begin{align}
\textbf{W}=\frac{d}{c(c+d)}\left(\sigma_r^2\textbf{H}_e\textbf{H}^H_e+\textbf{h}_b\textbf{h}^H_b\right),~
\textbf{u}=\frac{\textbf{h}_b\textbf{h}^H_b\widehat{\boldsymbol{\theta}}}{c}.
\end{align}
Then, the problem (\ref{SNR3}) can be reformulated as
\begin{subequations}\label{SNR33}
\begin{align}
&\min \limits_{\widehat{\boldsymbol{\theta}}}
~~\widehat{\boldsymbol{\theta}}^H\textbf{W}\widehat{\boldsymbol{\theta}}-2\Re\{\hat{\boldsymbol{\theta}}^H\textbf{u}\}\\
&~~\text{s.t.} ~~~|\widehat{\boldsymbol{\theta}}(m)|=1, \text{if}~ m\not\in \Omega,\\
& ~~~~~~~~|\widehat{\boldsymbol{\theta}}(m)|\leq \beta_{\text{max}}, \text{if}~m\in \Omega, \\
&~~~~~~~~|\widehat{\boldsymbol{\theta}}(m+1)|=1, ~(\ref{Pr2}).
\end{align}
\end{subequations}
In accordance with the MM algorithm in \cite{Sun2017Majorization}, we have
\begin{align}
\textbf{x}^H\textbf{L}\textbf{x}\leq \textbf{x}^H\textbf{T}\textbf{x}+
2\Re(\textbf{x}^H(\textbf{L}-\textbf{T})\textbf{x}_t)+\textbf{x}^H_t(\textbf{T}-\textbf{L})\textbf{x}_t,
\end{align}
where $\textbf{T}\succeq\textbf{L}$, and the equation holds when $\textbf{x}=\textbf{x}_t$. Then, the objective function in (\ref{SNR33}) can be recasted as
\begin{align}\label{widehate}
&\widehat{\boldsymbol{\theta}}^H\textbf{W}\widehat{\boldsymbol{\theta}}-
2\Re\{\widehat{\boldsymbol{\theta}}^H\textbf{u}\}\nonumber\\
&\leq \lambda_{\text{max}}(\textbf{W})\|\widehat{\boldsymbol{\theta}}\|^2-2\Re\{\widehat{\boldsymbol{\theta}}^H
(\lambda_{\text{max}}(\textbf{W})\textbf{I}-\textbf{W})\bar{\boldsymbol{\theta}}\}+\nonumber\\
&~~~\bar{\boldsymbol{\theta}}^H(\lambda_{\text{max}}\textbf{I}-\textbf{W})\widehat{\boldsymbol{\theta}}^{(i)}-
2\Re \{\widehat{\boldsymbol{\theta}}^H\textbf{u}\}\nonumber\\
&\leq \lambda_{\text{max}}(\textbf{W})(M_a\beta^2_{\text{max}}+M_p+1)+
\bar{\boldsymbol{\theta}}^H(\lambda_{\text{max}}(\textbf{W})\textbf{I}-\textbf{W})\bar{\boldsymbol{\theta}}\nonumber\\
&~~~-2\Re \{\widehat{\boldsymbol{\theta}}^H((\lambda_{\text{max}}(\textbf{W})\textbf{I}
-\textbf{W})\bar{\boldsymbol{\theta}}+\textbf{u})\}.
\end{align}
In accordance with (\ref{widehate}) and neglecting the constant term, the optimization problem (\ref{SNR33}) can be simplified to
\begin{subequations}\label{SNR4}
\begin{align}
&~\max \limits_{\widehat{\boldsymbol{\theta}}}
~~\Re \{\widehat{\boldsymbol{\theta}}^H
((\lambda_{\text{max}}(\textbf{W})\textbf{I}-\textbf{W})\bar{\boldsymbol{\theta}}+\textbf{u})\}\\
&~~\text{s.t.} ~~~|\widehat{\boldsymbol{\theta}}(m)|=1, \text{if}~ m\not\in \Omega,\\
& ~~~~~~~~|\widehat{\boldsymbol{\theta}}(m)|\leq \beta_{\text{max}}, \text{if}~m\in \Omega, \\
&~~~~~~~~|\widehat{\boldsymbol{\theta}}(m+1)|=1, ~(\ref{Pr2}).
\end{align}
\end{subequations}
It can be addressed by using optimization toolbox, such as CVX. The whole procedure of the Max-SNR-MM scheme is described in Algorithm 2.

\begin{algorithm}
\caption{Proposed Max-SNR-MM algorithm}
\begin{algorithmic}[1]
\STATE Initialize feasible solutions $\textbf{v}^{(0)}$ and $\boldsymbol{\Theta}^{(0)}$, calculate achievable rate $R^{(0)}_b$ based on (\ref{R_b}).
\STATE Set the iteration number $k=0$, accuracy value $\epsilon$.
\REPEAT
\STATE Given $\boldsymbol{\Theta}^{(k)}$, solve (\ref{v31}) to get $\textbf{v}^{(k+1)}$,.
\STATE Given $\textbf{v}^{(k+1)}$, solve (\ref{SNR4}) to get $\widehat{\boldsymbol{\theta}}^{(k+1)}$, and $\boldsymbol{\Theta}^{(k+1)}=\text{diag}(\widehat{\boldsymbol{\theta}}^{(k+1)}(1:M))^*$.
\STATE Calculate $R^{(k+1)}_b$ based on $\textbf{v}^{(k+1)}$ and $\boldsymbol{\Theta}^{(k+1)}$.
\STATE Update $k=k+1$.
\UNTIL {$|R_b^{(k)}-R_b^{(k-1)}|\leq\epsilon$.}
\end{algorithmic}
\end{algorithm}

The overall computational complexity of the proposed Max-SNR-MM algorithm is $\mathcal {O}(G(4(M+1)^2+N^3+3N^2+4N^2M+8MN)$ FLOPs, where $G$ stands for the numbers of alternating iterations, $\epsilon$ denotes the accuracy.

\section{Simulation Results}\label{s5}
Simulation results are shown to examine the performance of three proposed schemes in this section. Simulation parameters are given as follows: $N=8$, $M=128$, $M_a=32$, $d=\lambda/2$,
%$\theta_{sr}=\pi/4$, $\theta_{sb}=\pi/3$, $d_{sr}=200$m, $d_{sb}=220$m,
$\sigma^2_b=-70$dBm, $\sigma^2_r=2\sigma^2_b$, $P=25$dBm, $P_r^{\text{max}}=30$dBm. The base station, IRS (or UAV), and Bob are located at (0m, 0m, 0m), ($100\sqrt{2}$m, 0m, $100\sqrt{2}$m), ($110\sqrt{3}$m, 0m, 110m), respectively. The PL at the distance $d_{ab}$ is given by $\rho(d_{ab})=\text{PL}_0-10\gamma\text{log}_{10}\frac{d_{ab}}{d_0}$, where $\gamma$ represents the PL exponent, and $\text{PL}_0=-30$dB represents the PL reference distance $d_0=1$m. The PL exponents of all channels are chosen as 2. We select the positions of the hybrid IRS active elements as $\Omega=\{1, \cdots, M_a\}$.

%\begin{figure*}
%	\setlength{\abovecaptionskip}{-5pt}
%	\setlength{\belowcaptionskip}{-10pt}
%	\centering
%	\begin{minipage}[t]{0.33\linewidth}
%		\centering
%		\includegraphics[width=2.56in]{itea_new.eps}
%		\caption{Convergence of the proposed algorithms.}\label{itea}
%	\end{minipage}%
%	\begin{minipage}[t]{0.33\linewidth}
%		\centering
%		\includegraphics[width=2.56in]{complex_new.eps}
%		\caption{Computational complexity versus the numbers of RIS elements.}\label{complex}
%	\end{minipage}
%	\begin{minipage}[t]{0.33\linewidth}
%		\centering
%		\includegraphics[width=2.56in]{AR_M_new.eps}
%		\caption{Achievable rate versus the numbers of RIS phase shift elements.}\label{AR_m}
%	\end{minipage}
%\end{figure*}

\begin{figure}[htbp]
\centering
\includegraphics[width=0.5\textwidth]{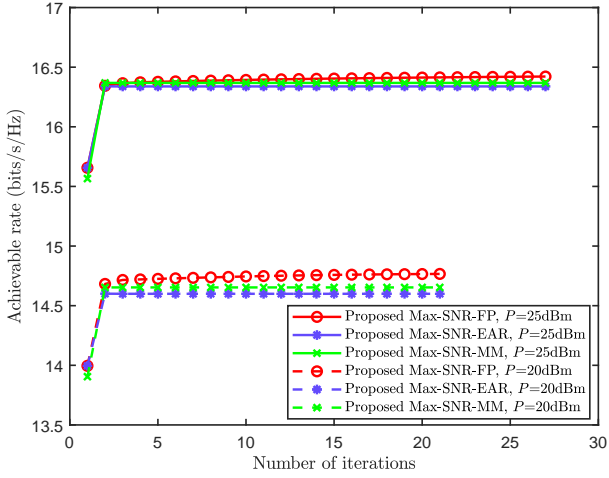}\\
\caption{Convergence of the proposed methods at different base station power.}\label{itea}
\end{figure}

Firstly, the convergence behaviour of the three proposed algorithms, called Max-SNR-FP, Max-SNR-EAR, and Max-SNR-MM, is investigated. Fig.~\ref{itea} presents the achievable rate versus the different base station power, i.e., $P=20$dBm, 25dBm. From Fig.~\ref{itea}, it can be seen that all of the proposed methods converge taking only a finite number of iterations. The proposed Max-SNR-EAR and Max-SNR-MM methods have a faster convergence rate than the Max-SNR-FP method, regardless of $P=20$dBm or 25dBm.
	
\begin{figure}[htbp]
\centering
\includegraphics[width=0.5\textwidth]{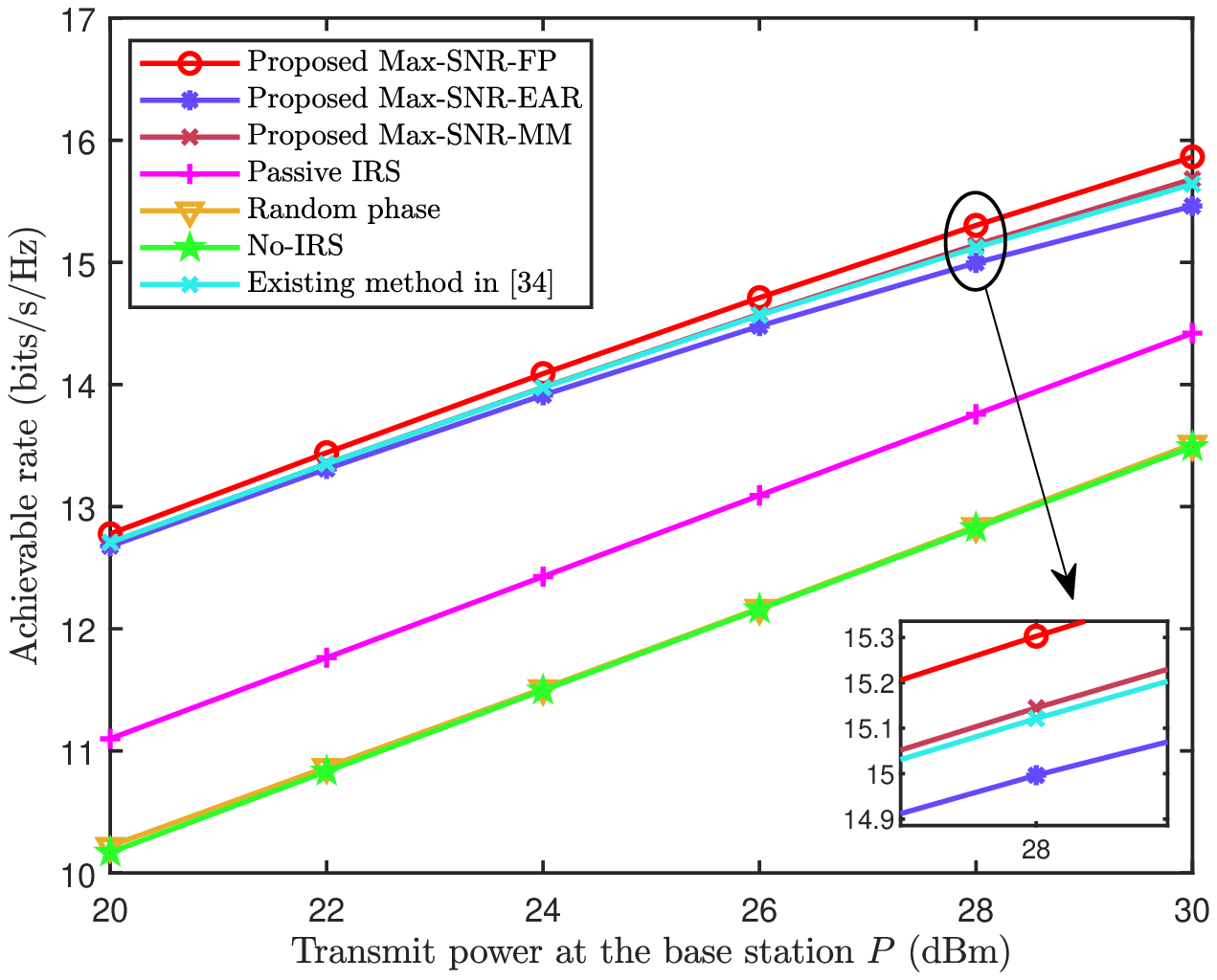}\\
\caption{Achievable rate versus the numbers of IRS phase shift elements.}\label{AR_P}
\end{figure}

Fig.~\ref{AR_P} demonstrates the curves of the achievable rate versus the base station transmit power $P$, where $M=32$ and $M_a=10$. As we can see from the figure that the achievable rates of three proposed methods and benchmark schemes increase gradually increases of $P$, and the achievable rates of the three proposed methods and existing method in \cite{Nguyen2022Hybrid2} are superior to those of other benchmark methods: passive IRS, random phase, and no IRS. Their achievable rates are about 15\% better than the passive IRS and about 25\% better than both the no-IRS and random phase schemes when $P= 20$. In addition, the achievable rate of the proposed Max-SNR-FP method is outperforms the proposed Max-SNR-MM method, the existing method in \cite{Nguyen2022Hybrid2}, and the proposed Max-SNR-EAR method regardless of the value of $P$.

\begin{figure}[htbp]
\centering
\includegraphics[width=0.5\textwidth]{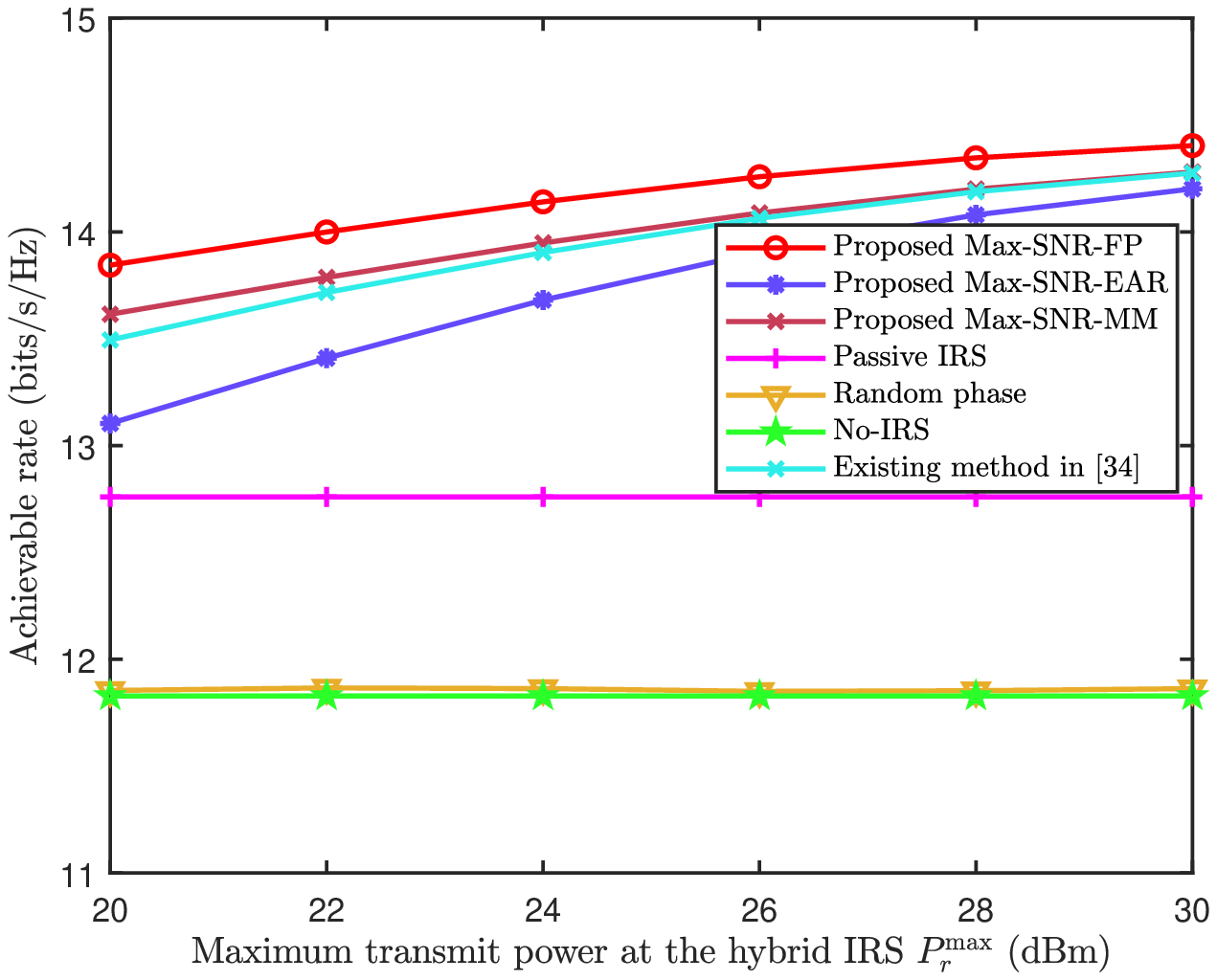}\\
\caption{Achievable rate versus the numbers of IRS phase shift elements.}\label{AR_Pr}
\end{figure}

Fig.~\ref{AR_Pr} shows the curves of the achievable rate versus the IRS power budget $P^{\max}_r$ ranging from 20dBm to 30dBm, where $M=32$ and $M_a=10$. The achievable rates of three proposed algorithms and existing algorithm in \cite{Nguyen2022Hybrid2} increase with the increases of the maximum transmit power $P^{\max}_r$. This is due to the fact that the hybrid IRS with active elements provides more performance gain with the increasing $P^{\max}_r$. As $P^{\max}_r$ increases, the difference among the achievable rates of the proposed Max-SNR-FP scheme, the proposed Max-SNR-EAR scheme, the proposed Max-SNR-MM scheme, and the existing scheme in \cite{Nguyen2022Hybrid2} gradually decreases. The decreasing order of their achievable rates is Max-SNR-FP, Max-SNR-MM, the existing scheme in \cite{Nguyen2022Hybrid2}, and Max-SNR-EAR.

\begin{figure}[htbp]
\centering
\includegraphics[width=0.5\textwidth]{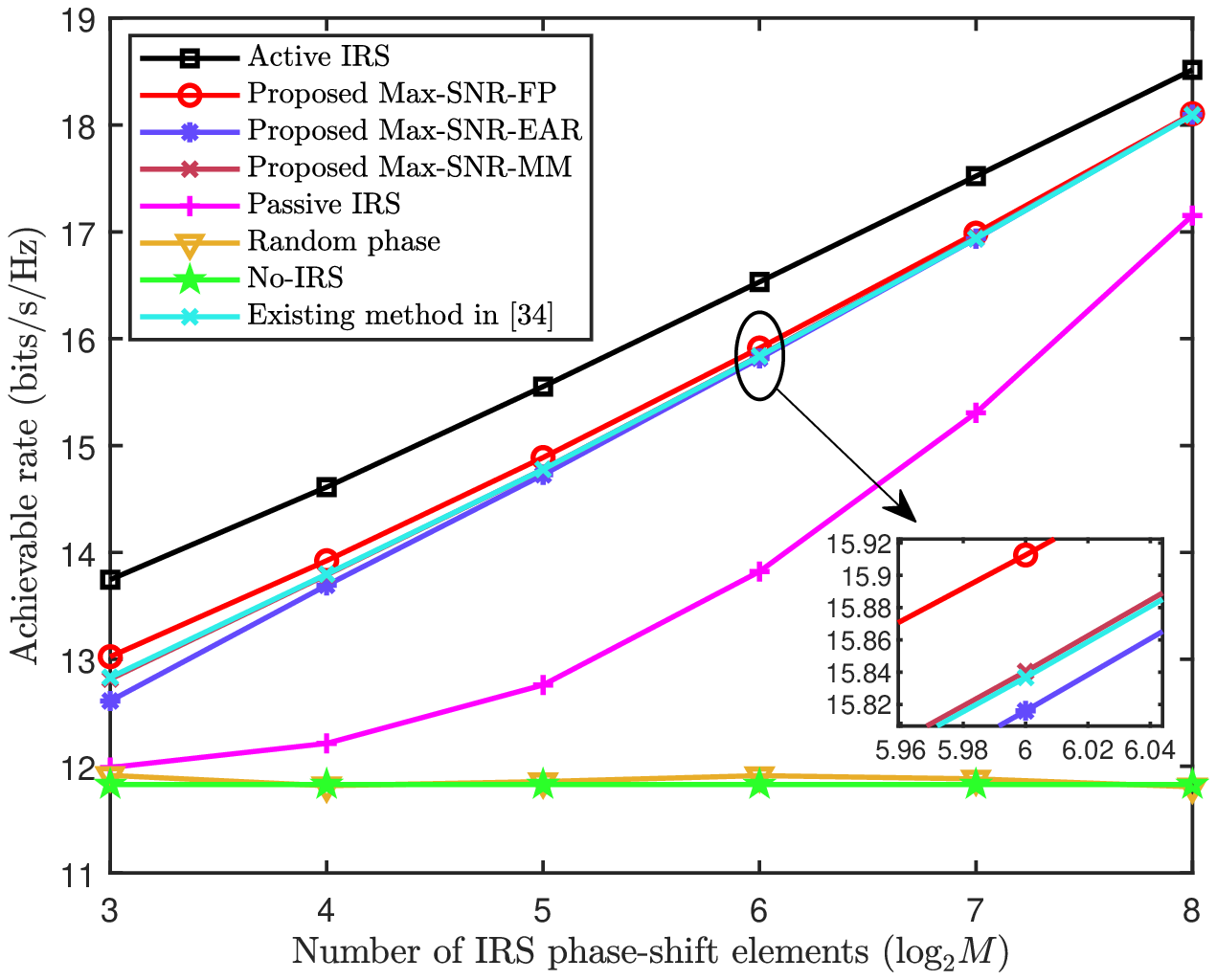}\\
\caption{Achievable rate versus the numbers of IRS phase shift elements.}\label{AR_m}
\end{figure}

Fig.~\ref{AR_m} depicts the curves of the achievable rate versus the number $M$ of hybrid IRS phase shift elements, where $M=2M_a$. We compare the achievable rates of three proposed methods with those of the benchmark schemes: active IRS, passive IRS, no IRS, random phase IRS, and existing method in \cite{Nguyen2022Hybrid2}. The achievable rates of the proposed Max-SNR-FP, Max-SNR-EAR, and Max-SNR-MM schemes gradually increase with the increases of $M$, and the first proposed sheme is better than the rest and existing method in \cite{Nguyen2022Hybrid2}. The achievable rates of three proposed schemes outperform those of the passive IRS, random phase IRS, and without IRS. In addition, when $M$ tends to large-scale, the difference in achievable rates between the three proposed schemes and active IRS gradually decreases. %These further validates the advantages of using a hybrid RIS.

\begin{figure}[htbp]
\centering
\includegraphics[width=0.5\textwidth]{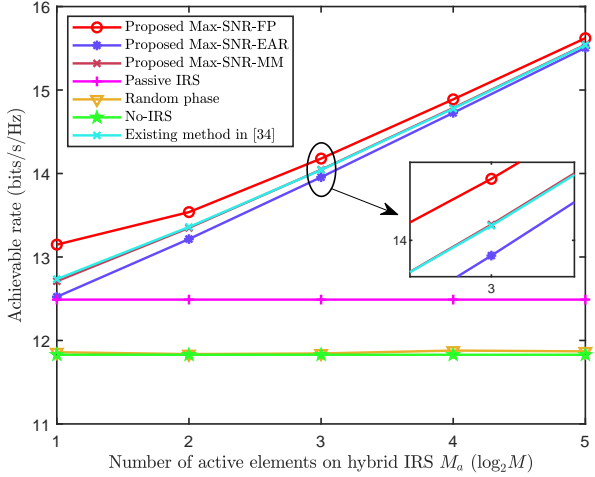}\\
\caption{Achievable rate versus the numbers of IRS phase shift elements.}\label{AR_Ma}
\end{figure}

Fig.~\ref{AR_Ma} presents the achievable rate versus the number $M_a$ of active elements on hybrid IRS of three proposed schemes and benchmark schemes. It is observed that when $M=2$, the achievable rate of the proposed Max-SNR-EAR method is similar to that of the passive IRS. The difference of the achievable rates between the proposed Max-SNR-MM method and the existing method in \cite{Nguyen2022Hybrid2} method is trivial, regardless of the value of $M_a$. With the increase of $M_a$, the hybrid IRS is converted to active IRS, and the achievable rates of three proposed methods and the existing method in \cite{Nguyen2022Hybrid2} gradually increase, while that of passive IRS, random phase IRS, and no IRS is maintained. This illustrates the advantages of active elements in hybrid IRS for performance enhancement of the communication networks.

\begin{figure}[htbp]
\centering
\includegraphics[width=0.5\textwidth]{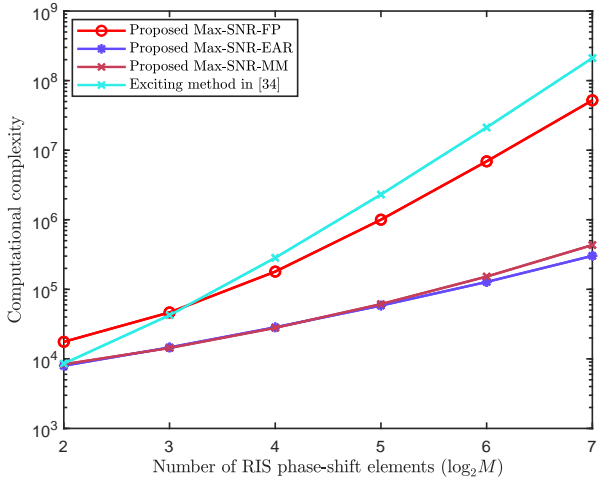}\\
\caption{Computational complexity versus the numbers of IRS elements.}\label{complex}
\end{figure}

Fig.~\ref{complex} plots the curves of the computational complexity versus the number $M$ of hybrid IRS elements. It can be found that the difference between the computational complexity of the proposed Max-SNR-EAR and Max-SNR-MM methods increases with the increase of M. Moreover, when $M$ tends to large-scale, the computational complexities of the existing method in \cite{Nguyen2022Hybrid2} and proposed Max-SNR-FP method are far higher than those of the proposed Max-SNR-MM and proposed Max-SNR-EAR methods.

\section{Conclusion}\label{s6}
In this work,  the hybrid-IRS-and-UAV-assisted DM network was investigated. To fully explore the advantages of the hybrid IRS and maximize the achievable rate, the Max-SNR-FP, Max-SNR-EAR, and Max-SNR-MM methods were proposed to jointly design the beamforming vector, passive IRS PSM, and active IRS PSM by alternately optimizing one and giving rest. From the simulation results, it can be found that the achievable rate of three proposed methods increases with the number of hybrid IRS elements increases, and is superior to those of without IRS, random phase IRS, and passive IRS. The difference of the achievable rates between the proposed Max-SNR-MM method and the existing method in \cite{Nguyen2022Hybrid2} is trivial.
%regardless of the values of BS power, maximum transmit power at IRS, the number of the IRS elements, and the number of
%active elements on IRS.
%Moreover, the proposed Max-SNR-FP and Max-SNR-MM methods exceeds the existing method with respect to the achievable rate and has lower computational complexity.
With the increase of the number of IRS elements, the decreasing order of their achievable rates is Max-SNR-FP, Max-SNR-MM, the existing scheme in \cite{Nguyen2022Hybrid2}, and Max-SNR-EAR, while the decreasing order of the computational complexity is existing scheme in \cite{Nguyen2022Hybrid2}, Max-SNR-FP, Max-SNR-MM, and Max-SNR-EAR.
%\section{Numerical results}

%\section{Conclusions}

\ifCLASSOPTIONcaptionsoff
  \newpage
\fi

\bibliographystyle{IEEEtran}

\bibliography{IEEEfull,reference}

\begin{thebibliography}{10}
\providecommand{\url}[1]{#1}
\csname url@samestyle\endcsname
\providecommand{\newblock}{\relax}
\providecommand{\bibinfo}[2]{#2}
\providecommand{\BIBentrySTDinterwordspacing}{\spaceskip=0pt\relax}
\providecommand{\BIBentryALTinterwordstretchfactor}{4}
\providecommand{\BIBentryALTinterwordspacing}{\spaceskip=\fontdimen2\font plus
\BIBentryALTinterwordstretchfactor\fontdimen3\font minus
  \fontdimen4\font\relax}
\providecommand{\BIBforeignlanguage}[2]{{%
\expandafter\ifx\csname l@#1\endcsname\relax
\typeout{** WARNING: IEEEtran.bst: No hyphenation pattern has been}%
\typeout{** loaded for the language `#1'. Using the pattern for}%
\typeout{** the default language instead.}%
\else
\language=\csname l@#1\endcsname
\fi
#2}}
\providecommand{\BIBdecl}{\relax}
\BIBdecl

\bibitem{Wu2018A}
Y.~Wu, A.~Khisti, C.~Xiao, G.~Caire, K.-K. Wong, and X.~Gao, ``A survey of
  physical layer security techniques for {5G} wireless networks and challenges
  ahead,'' \emph{IEEE J. Sel. Areas Commun.}, vol.~36, no.~4, pp. 679--695,
  Oct. 2018.

\bibitem{Zheng2013Improving}
G.~Zheng, I.~Krikidis, J.~Li, A.~P. Petropulu, and B.~Ottersten, ``Improving
  physical layer secrecy using full-duplex jamming receivers,'' \emph{IEEE
  Trans. Signal Process.}, vol.~61, no.~20, pp. 4962--4974, Jun. 2013.

\bibitem{Zeng2016Wireless}
Y.~Zeng, R.~Zhang, and T.~J. Lim, ``Wireless communications with unmanned
  aerial vehicles: opportunities and challenges,'' \emph{IEEE Commun. Mag.},
  vol.~54, no.~5, pp. 36--42, May. 2016.

\bibitem{Yin2020Resource}
S.~Yin, L.~Li, and F.~R. Yu, ``Resource allocation and basestation placement in
  downlink cellular networks assisted by multiple wireless powered {UAVs},''
  \emph{IEEE Trans. Veh. Technol.}, vol.~69, no.~2, pp. 2171--2184, Feb. 2020.

\bibitem{Zhang2018Joint}
S.~Zhang, H.~Zhang, Q.~He, K.~Bian, and L.~Song, ``Joint trajectory and power
  optimization for {UAV} relay networks,'' \emph{IEEE Commun. Lett.}, vol.~22,
  no.~1, pp. 161--164, Jan. 2018.

\bibitem{Zhou2021UAV}
X.~Zhou, S.~Yan, F.~Shu, R.~Chen, and J.~Li, ``{UAV}-enabled covert wireless
  data collection,'' \emph{IEEE J. Sel. Areas Commun.}, vol.~39, no.~11, pp.
  3348--3362, Nov. 2021.

\bibitem{Wen20233D}
F.~Wen, J.~Shi, G.~Gui, H.~Gacanin, and O.~A. Dobre, ``{3-D} positioning method
  for anonymous {UAV} based on bistatic polarized {MIMO} radar,'' \emph{IEEE
  Internet Things J.}, vol.~10, no.~1, pp. 815--827, Jan. 2023.

\bibitem{Azari2020UAV}
M.~M. Azari, G.~Geraci, A.~Garcia-Rodriguez, and S.~Pollin, ``{UAV}-to-{UAV}
  communications in cellular networks,'' \emph{IEEE Trans. Wirel. Commun.},
  vol.~19, no.~9, pp. 6130--6144, Sep. 2020.

\bibitem{Zeng2021Trajectory}
S.~Zeng, H.~Zhang, B.~Di, and L.~Song, ``Trajectory optimization and resource
  allocation for {OFDMA} {UAV} relay networks,'' \emph{IEEE Trans. Wirel.
  Commun.}, vol.~20, no.~10, pp. 6634--6647, Oct. 2021.

\bibitem{Wu2021UAV}
Y.~Wu, W.~Yang, X.~Guan, and Q.~Wu, ``{UAV}-enabled relay communication under
  malicious jamming: Joint trajectory and transmit power optimization,''
  \emph{IEEE Trans. Veh. Technol.}, vol.~70, no.~8, pp. 8275--8279, Aug. 2021.

\bibitem{Wu2019Intelligent}
Q.~Wu and R.~Zhang, ``Intelligent reflecting surface enhanced wireless network
  via joint active and passive beamforming,'' \emph{IEEE Trans. Wirel.
  Commun.}, vol.~18, no.~11, pp. 5394--5409, Nov. 2019.

\bibitem{Pan2021UAV}
Y.~Pan, C.~Wang, C.~Pan, H.~Zhu, and J.~Wang, ``{UAV}-assisted and intelligent
  reflecting surfaces-supported terahertz communication,'' \emph{Wireless
  Commun. Lett.}, vol.~10, no.~6, pp. 1256--1260, Jun. 2021.

\bibitem{Su2022Spectrum}
Y.~Su, X.~Pang, S.~Chen, X.~Jiang, N.~Zhao, and F.~R. Yu, ``Spectrum and energy
  efficiency optimization in irs-assisted uav networks,'' \emph{IEEE Trans
  Commun.}, vol.~70, no.~10, pp. 6489--6502, Oct. 2022.

\bibitem{Fang2021Joint}
S.~Fang, G.~Chen, and Y.~Li, ``Joint optimization for secure intelligent
  reflecting surface assisted {UAV} networks,'' \emph{IEEE Wireless Commun.
  Lett.}, vol.~10, no.~2, pp. 276--280, Feb. 2021.

\bibitem{Pang2022IRS}
X.~Pang, N.~Zhao, J.~Tang, C.~Wu, D.~Niyato, and K.-K. Wong, ``{IRS}-assisted
  secure {UAV} transmission via joint trajectory and beamforming design,''
  \emph{IEEE Trans Commun.}, vol.~70, no.~2, pp. 1140--1152, Feb. 2022.

\bibitem{Hua2021UAV}
M.~Hua, L.~Yang, Q.~Wu, C.~Pan, C.~Li, and A.~L. Swindlehurst, ``{UAV}-assisted
  intelligent reflecting surface symbiotic radio system,'' \emph{IEEE
  Transactions on Wireless Communications}, vol.~20, no.~9, pp. 5769--5785,
  Sep. 2021.

\bibitem{Cheng2021Physical}
Q.~Cheng, S.~Wang, V.~Fusco, F.~Wnag, J.~Zhu, and C.~Gu, ``Physical-layer
  security for frequency diverse array-based directional modulation in
  fluctuating two-ray fading channels,'' \emph{IEEE Trans. Wirel. Commun.},
  vol.~20, no.~7, pp. 4190--4204, Jul. 2021.

\bibitem{Wang2018Hybrid}
W.-Q. Wang and Z.~Zheng, ``Hybrid {MIMO} and phased-array directional
  modulation for physical layer security in mmwave wireless communications,''
  \emph{IEEE J. Sel. Areas Commun.}, vol.~36, no.~7, pp. 1383--1396, Jul. 2018.

\bibitem{Nusenu2019Development}
S.~Y. Nusenu, ``Development of frequency modulated array antennas for
  millimeter-wave communications,'' \emph{Wireless Commun. Mobile Comput.},
  vol. 2019, pp. 1--16, Apr. 2019.

\bibitem{Qiu2020Multi}
B.~Qiu, L.~Wang, J.~Xie, Z.~Zhang, Y.~Wang, and M.~Tao, ``Multi{-}beam index
  modulation with cooperative legitimate users schemes based on frequency
  diverse array,'' \emph{IEEE Trans. Veh. Technol.}, vol.~69, no.~10, pp.
  11\,028--11\,041, Oct. 2020.

\bibitem{Daly2009Directional}
M.~P. Daly and J.~T. Bemhard, ``Directional modulation technique for phased
  arrays,'' \emph{IEEE Trans. Antennas Propag}, vol.~57, no.~9, pp. 2633--2640,
  Sep. 2009.

\bibitem{Daly2010Beamsteering}
M.~P. Daly and J.~T. Bernhard, ``Beamsteering in pattern reconfigurable arrays
  using directional modulation,'' \emph{IEEE T Antenn Propag}, vol.~58, no.~7,
  pp. 2259--2265, Mar. 2010.

\bibitem{Shu2016Robust}
F.~Shu, X.~Wu, J.~Li, R.~Chen, and B.~Vucetic, ``Robust synthesis scheme for
  secure multi-beam directional modulation in broadcasting systems,''
  \emph{IEEE Access}, vol.~4, pp. 6614--6623, Nov. 2016.

\bibitem{Xie2018Artificial}
T.~Xie, J.~Zhu, and Y.~Li, ``Artificial-noise-aided zero-forcing synthesis
  approach for secure multi-beam directional modulation,'' \emph{IEEE Commun
  Lett.}, vol.~22, no.~2, pp. 276--279, Feb. 2018.

\bibitem{Teng2022Low}
Y.~Teng, J.~Li, M.~Huang, L.~Liu, G.~Xia, X.~Zhou, F.~Shu, and J.~Wang,
  ``Low-complexity and high-performance receive beamforming for secure
  directional modulaion networks against an eavesdropping-enabled full-duplex
  attacker,'' \emph{Sci China Inf. Sci.}, vol.~65, pp. 119\,302--119\,302, Jan.
  2022.

\bibitem{ShuEnhanced2021}
F.~Shu, Y.~Teng, J.~Li, M.~Huang, W.~Shi, J.~Li, Y.~Wu, and J.~Wang, ``Enhanced
  secrecy rate maximization for directional modulation networks via {IRS},''
  \emph{IEEE Trans. Commun.}, vol.~69, no.~12, pp. 8388--8401, Dec. 2021.

\bibitem{Dong2022Low}
R.~Dong, S.~Jiang, X.~Hua, Y.~Teng, F.~Shu, and J.~Wang, ``Low-complexity joint
  phase adjustment and receive beamforming for directional modulation networks
  via {IRS},'' \emph{IEEE open journal of the Communications Society}, vol.~3,
  pp. 1234--1243, Aug. 2022.

\bibitem{Chen2022Artificial}
J.~Chen, Y.~Xiao, X.~Lei, H.~Niu, and Y.~Yuan, ``Artificial noise aided
  directional modulation via reconfigurable intelligent surface: Secrecy
  guarantee in range domain,'' \emph{IET Commun.}, vol.~16, pp. 1558--1569,
  Mar. 2022.

\bibitem{Zhang2021Active}
Z.~Zhang, L.~Dai, X.~Chen, C.~Liu, F.~Yang, R.~Schober, and H.~V. Poor,
  ``Active {RIS} vs. passive {RIS}: which will previal in {6G}?'' \emph{IEEE
  Trans Commun.}, pp. 1--1, 2022.

\bibitem{Liu2022Active}
K.~Liu, Z.~Zhang, L.~Dai, s.~Xu, and F.~Yang, ``Active reconfigurable
  intelligent surface: Fully-connected or sub-connected?'' \emph{IEEE Commun.
  Lett.}, vol.~26, no.~1, pp. 167--171, Jan. 2022.

\bibitem{Ren2023Transmission}
H.~Ren, Z.~Chen, G.~Hu, Z.~Peng, C.~Pan, and J.~Wang, ``Transmission design for
  active {RIS}-aided simultaneous wireless information and power transfer,''
  \emph{IEEE Wireless Commun. Lett.}, pp. 1--1, 2023.

\bibitem{Dong2022Active}
L.~Dong, H.-M. Wang, and J.~Bai, ``Active reconfigurable intelligent surface
  aided secure transmission,'' \emph{IEEE Trans. Veh. Technol.}, vol.~71,
  no.~2, pp. 2181--2186, Dec. 2022.

\bibitem{Lv2023RIS}
W.~Lv, J.~Bai, Q.~Yan, and H.~M. Wang, ``Ris-assisted green secure
  communications: Active {RIS} or passive {RIS}?'' \emph{IEEE Wireless Commun.
  Lett}, vol.~12, no.~2, pp. 237--241, Feb. 2023.

\bibitem{Nguyen2022Hybrid2}
N.~T. Nguyen, V.-D. Nguyen, Q.~Wu, A.~T\"{o}lli, S.~Chatzinotas, and M.~Juntti,
  ``Hybrid active-passive reconfigurable intelligent surface-assisted
  multi-user {MISO} systems,'' \emph{2022 IEEE 23rd International Workshop on
  Signal Processing Advances in Wireless Communication (SPAWC)}, pp. 1--5, Jul.
  2022.

\bibitem{Nguyen2022Hybrid}
N.~T. Nguyen, Q.-D. Vu, K.~Lee, and M.~Juntti, ``Hybrid relay-reflecting
  intelligent surface-assisted wireless communications,'' \emph{IEEE Trans.
  Veh. Technol.}, Mar. 2022.

\bibitem{Nguyen2021Spectral}
------, ``Spectral efficiency optimization for hybrid relay-reflecting
  intelligent surface,'' \emph{2021 IEEE International Conference on
  Communications Workshops (ICC Workshops)}, pp. 1--6, 2021.

\bibitem{Ngo2021Low}
K.-H. Ngo, N.~T. Nguyen, T.~Q. Dinh, T.-M. Hoang, and M.~Juntti, ``Low-latency
  and secure computation offloading assisted by hybrid relay-reflecting
  intelligent surface,'' \emph{2021 International Conference on Advanced
  Technologies for Communications (ATC)}, pp. 306--311, 2021.

\bibitem{Hu2021Hybrid}
J.~Hu, X.~Shi, S.~Yan, Y.~Chen, T.~Zhao, and F.~Shu, ``Hybrid relay-reflecting
  intelligent surface-aided covert communications,''
  \emph{https://arxiv.org/abs/2203.12223}.

\bibitem{Sankar2022Beamforming}
R.~P. Sankar and S.~P. Chepuri, ``Beamforming in hybrid {RIS} assisted
  integrated sensing and communication systems,'' \emph{2022 30th European
  Signal Processing Conference (EUSIPCO)}, pp. 1082--1086, Aug. 2022.

\bibitem{Pan2020Multicell}
C.~Pan, H.~Ren, K.~Wang, W.~Xu, M.~Elkashlan, A.~Nallanathan, and L.~Hanzo,
  ``Multicell {MIMO} communications relaying on intelligent reflecting
  surfaces,'' \emph{IEEE Trans. Wirel. Commun.}, vol.~19, no.~8, pp.
  5218--5233, Aug. 2020.

\bibitem{Wang2020Channel}
Z.~Wang, L.~Liu, and S.~Cui, ``Channel estimation for intelligent reflecting
  surface assisted multiuser communications: Framework, algorithms, and
  analysis,'' \emph{IEEE Trans. Wirel. Commun.}, vol.~19, no.~10, pp.
  6607--6620, Oct. 2020.

\bibitem{Shi2021Secure}
W.~Shi, J.~Li, G.~Xia, Y.~Wang, X.~Zhou, Y.~Zhang, and F.~Shu, ``Secure
  multigroup multicast communication systems via intelligent reflecting
  surface,'' \emph{China Commun.}, vol.~18, no.~3, pp. 39--51, Mar. 2021.

\bibitem{Dinkelbach1967On}
W.~Dinkelbach, ``On nonlinear fractional programming,'' \emph{Manage Sci.},
  vol.~13, no.~7, pp. 492--498, Mar. 1967.

\bibitem{Sadek2007A}
M.~Sadek, A.~Tarighat, and A.~H. Sayed, ``A leakage-based precoding scheme for
  downlink multi-user {MIMO} channels,'' \emph{IEEE Trans. Wirel. Commun.},
  vol.~6, no.~5, pp. 1711--1721, May. 2007.

\bibitem{Nasir2017Secrecy}
A.~A. Nasir, H.~D. Tuan, T.~Q. Duong, and H.~V. Poor, ``Secrecy rate
  beamforming for multicell networks with information and energy harvesting,''
  \emph{IEEE Trans. Signal Process.}, vol.~65, no.~3, pp. 677--689, Oct. 2017.

\bibitem{Sun2017Majorization}
Y.~Sun, P.~Babu, and D.~P. Palomar, ``Majorization-minimization algorithms in
  signal processing, communications, and machine learning,'' \emph{IEEE Trans.
  Signal Process.}, vol.~65, no.~3, pp. 794--816, Aug. 2017.

\end{thebibliography}
\end{document}